\documentclass[envcountsame]{llncs}
\usepackage{amssymb}

\usepackage{times}

\title{Topological Complexity of Context-Free $\omega$-Languages : \\
A Survey}
\author{Olivier Finkel}
\institute{{\it Equipe de Logique Math\'ematique}  
 \\  Institut de Math\'ematiques de Jussieu - Paris Rive Gauche
                                                   UMR7586
       \\  CNRS et Universit\'e Paris Diderot Paris 7
\\     B\^{a}timent Sophie Germain                       
           Case 7012                                                    
                     \\75205 Paris Cedex 13, France. \\ \email{finkel@math.univ-paris-diderot.fr} }

\date{}
\begin{document}

\spnewtheorem{Rem}[theorem]{Remark}{\bfseries}{\itshape}
\spnewtheorem{Exa}[theorem]{Example}{\bfseries}{\itshape}
\spnewtheorem{The}[theorem]{Theorem}{\bfseries}{\itshape}

\spnewtheorem{Pro}[theorem]{Proposition}{\bfseries}{\itshape}
\spnewtheorem{lem}[theorem]{Lemma}{\bfseries}{\itshape}
\spnewtheorem{cor}[theorem]{Corollary}{\bfseries}{\itshape}
\spnewtheorem{Deff}[theorem]{Definition}{\bfseries}{\itshape}

\def\ufootnote#1{\let\savedthfn\thefootnote\let\thefootnote\relax
\footnote{#1}\let\thefootnote\savedthfn\addtocounter{footnote}{-1}}

\newcommand{\bormxi}{{\bf\Pi}^{0}_{\xi}}
\newcommand{\bormlxi}{{\bf\Pi}^{0}_{<\xi}}
\newcommand{\bormz}{{\bf\Pi}^{0}_{0}}
\newcommand{\bormone}{{\bf\Pi}^{0}_{1}}
\newcommand{\ca}{{\bf\Pi}^{1}_{1}}
\newcommand{\bormtwo}{{\bf\Pi}^{0}_{2}}
\newcommand{\bormthree}{{\bf\Pi}^{0}_{3}}
\newcommand{\bormom}{{\bf\Pi}^{0}_{\omega}}
\newcommand{\borom}{{\bf\Delta}^{0}_{\omega}}
\newcommand{\borml}{{\bf\Pi}^{0}_{\lambda}}
\newcommand{\bormlpn}{{\bf\Pi}^{0}_{\lambda +n}}
\newcommand{\bormpm}{{\bf\Pi}^{0}_{1+m}}
\newcommand{\borapm}{{\bf\Sigma}^{0}_{1+m}}
\newcommand{\bormep}{{\bf\Pi}^{0}_{\eta +1}}
\newcommand{\borapxi}{{\bf\Sigma}^{0}_{\xi}}
\newcommand{\borai}{{\bf\Sigma}^{0}_{ 2.\xi +1 }}
\newcommand{\bormpxi}{{\bf\Pi}^{0}_{\xi}}
\newcommand{\bormpeta}{{\bf\Pi}^{0}_{1+\eta}}
\newcommand{\borapxipo}{{\bf\Sigma}^{0}_{\xi +1}}
\newcommand{\bormpxipo}{{\bf\Pi}^{0}_{\xi +1}}
\newcommand{\borpxi}{{\bf\Delta}^{0}_{\xi}}
\newcommand{\borel}{{\bf\Delta}^{1}_{1}}
\newcommand{\Borel}{{\it\Delta}^{1}_{1}}
\newcommand{\borone}{{\bf\Delta}^{0}_{1}}
\newcommand{\bortwo}{{\bf\Delta}^{0}_{2}}
\newcommand{\borthree}{{\bf\Delta}^{0}_{3}}
\newcommand{\boraone}{{\bf\Sigma}^{0}_{1}}
\newcommand{\boratwo}{{\bf\Sigma}^{0}_{2}}
\newcommand{\borathree}{{\bf\Sigma}^{0}_{3}}
\newcommand{\boraom}{{\bf\Sigma}^{0}_{\omega}}
\newcommand{\boraxi}{{\bf\Sigma}^{0}_{\xi}}
\newcommand{\ana}{{\bf\Sigma}^{1}_{1}}
\newcommand{\pca}{{\bf\Sigma}^{1}_{2}}
\newcommand{\Ana}{{\it\Sigma}^{1}_{1}}
\newcommand{\Boraone}{{\it\Sigma}^{0}_{1}}
\newcommand{\Borone}{{\it\Delta}^{0}_{1}}
\newcommand{\Bormone}{{\it\Pi}^{0}_{1}}
\newcommand{\Bormtwo}{{\it\Pi}^{0}_{2}}
\newcommand{\Ca}{{\it\Pi}^{1}_{1}}
\newcommand{\bormn}{{\bf\Pi}^{0}_{n}}
\newcommand{\bormm}{{\bf\Pi}^{0}_{m}}
\newcommand{\boralp}{{\bf\Sigma}^{0}_{\lambda +1}}
\newcommand{\borat}{{\bf\Sigma}^{0}_{|\theta |}}
\newcommand{\bormat}{{\bf\Pi}^{0}_{|\theta |}}
\newcommand{\Borapxi}{{\it\Sigma}^{0}_{\xi}}
\newcommand{\Bormpxipo}{{\it\Pi}^{0}_{1+\xi +1}}
\newcommand{\Borapn}{{\it\Sigma}^{0}_{1+n}}
\newcommand{\borapn}{{\bf\Sigma}^{0}_{1+n}}
\newcommand{\boraxipm}{{\bf\Sigma}^{0}_{\xi^\pm}}
\newcommand{\Boratwo}{{\it\Sigma}^{0}_{2}}
\newcommand{\Borathree}{{\it\Sigma}^{0}_{3}}
\newcommand{\Borapnpo}{{\it\Sigma}^{0}_{1+n+1}}
\newcommand{\Bormpxi}{{\it\Pi}^{0}_{\xi}}
\newcommand{\Borpxi}{{\it\Delta}^{0}_{\xi}}
\newcommand{\boratpxi}{{\bf\Sigma}^{0}_{2+\xi}}
\newcommand{\Boratpxi}{{\it\Sigma}^{0}_{2+\xi}}
\newcommand{\bormltpxi}{{\bf\Pi}^{0}_{<2+\xi}}
\newcommand{\Bormltpxi}{{\it\Pi}^{0}_{<2+\xi}}
\newcommand{\borapeap}{{\bf\Sigma}^{0}_{1+\eta_{\alpha ,p}}}
\newcommand{\borapeapn}{{\bf\Sigma}^{0}_{1+\eta_{\alpha ,p,n}}}
\newcommand{\Borapeap}{{\it\Sigma}^{0}_{1+\eta_{\alpha ,p}}}
\newcommand{\Bormpn}{{\it\Pi}^{0}_{1+n}}
\newcommand{\Borpn}{{\it\Delta}^{0}_{1+n}}
\newcommand{\borapximo}{{\bf\Sigma}^{0}_{1+(\xi -1)}}
\newcommand{\borpeta}{{\bf\Delta}^{0}_{1+\eta}}

\newcommand{\hs}{\hspace{12mm}

\noi}
\newcommand{\noi}{\noindent}
\newcommand{\ol}{ $\omega$-language}
\newcommand{\om}{\omega}
\newcommand{\Si}{\Sigma}
\newcommand{\Sis}{\Sigma^\star}
\newcommand{\Sio}{\Sigma^\omega}

\newcommand{\nl}{\newline}
\newcommand{\lra}{\leftrightarrow}
\newcommand{\fa}{\forall}
\newcommand{\ra}{\rightarrow}
\newcommand{\orl}{ regular $\omega$-language}

\newcommand{\Ga}{\Gamma}
\newcommand{\Gas}{\Gamma^\star}

\newcommand{\la}{language}
\newcommand{\ite}{\item}
\newcommand{\Lp}{L(\varphi)}
\newcommand{\abs}{\{a, b\}^\star}
\newcommand{\abcs}{\{a, b, c \}^\star}

\newcommand{\tla}{\twoheadleftarrow}
\newcommand{\de}{deterministic }
\newcommand{\vp}{\varphi}
\newcommand{\proo}{\noi {\bf Proof.} }
\newcommand {\ep}{\hfill $\square$}

\maketitle

\begin{abstract}
We survey recent results on the topological complexity of context-free $\om$-languages which form the second level of the Chomsky hierarchy of 
languages of infinite words. In particular, we consider the Borel hierarchy and  the Wadge hierarchy of 
non-deterministic or deterministic context-free $\om$-languages. 
We study also decision problems, the  links with the notions of ambiguity and of  degrees of ambiguity, and the special case of $\om$-powers. 

\end{abstract}


\section{Introduction}

\noi The Chomsky hierarchy of formal  languages of finite words over a finite alphabet 
 is now well known,  \cite{HopcroftUllman79}. The class of regular languages accepted by finite automata 
forms the first level of this hierarchy and the class of context-free languages accepted by pushdown automata or generated by context-free grammars 
forms its second level \cite{ABB96}. The third and the fourth levels are formed by the class of context-sensitive languages accepted by linear-bounded automata 
or generated by Type-1 grammars and the class of recursively enumerable languages accepted by Turing machines or generated by Type-0 grammars
\cite{Chomsky}. 
In particular, context-free languages, firstly introduced by Chomsky to analyse the syntax of natural languages, 
have been very useful in  Computer Science,  in particular in the domain of programming languages, for the construction of compilers used to verify  
correctness of programs, 
\cite{HopcroftMotwaniUllman2001}. 

\hs There is a hierarchy of languages of infinite words which is  analogous to the Chomsky hierarchy but where the languages are formed by 
infinite words over a finite alphabet. The first level of this hierarchy is formed by the class of regular $\om$-languages 
 accepted by finite automata. They  were first studied by B\"uchi in order to study decision problems for logical theories. In particular,   B\"uchi
 proved that   the monadic second order theory of one successor
over the integers is decidable, using  finite automata equipped with a certain acceptance condition 
for infinite words, now called the     B\"uchi        acceptance condition. Well known  pioneers in this research area are named Muller,  Mc Naughton, 
Rabin, Landweber, Choueka, \cite{Muller63,Naughton66,Rabin69,Landweber69,Choueka74}.
The theory of  regular $\om$-languages is now well 
established and has found many applications for specification and verification 
of non-terminating systems;  see \cite{Thomas90,Staiger97,PerrinPin} for many results and references. 
 The second level of the hierarchy is formed by the class of context-free $\om$-languages. As in the case of languages of finite words it turned out that 
an $\om$-language is accepted by a (non-deterministic) pushdown automaton (with B\"uchi acceptance condition) if and only if it is generated by a context-free 
grammar where infinite derivations are considered. 
Context-free languages of infinite words were first studied by Cohen and Gold, \cite{CG,CG78},  Linna, \cite{Linna75,Linna76,Linna77}, Boasson, Nivat, 
\cite{Nivat78,Nivat77,Boasson79,BoassonNivat80}, Beauquier, \cite{Beauquier84}, see the survey \cite{Staiger97}. 
Notice that in the case of infinite words Type-1 grammars and Type-0 grammars accept the same $\om$-languages which are also the  $\om$-languages accepted by 
Turing machines with a B\"uchi acceptance condition \cite{CG78b,Staiger97}, see also  the fundamental study of   Engelfriet and  Hoogeboom on 
{\bf X}-automata, i.e. finite automata equipped with 
a storage type {\bf X}, accepting infinite words,\cite{eh}. 
\nl Context-free $\om$-languages have occurred recently in the works on games played on infinite pushdown graphs, following the fundamental study of 
Walukiewicz, \cite{Wal,Thomas02}  \cite{Serre04,Finkel05FI}.

\hs Since the set $X^\om$ of infinite words over a finite alphabet $X$ is naturally equipped with the Cantor topology, a 
way to  study  the complexity of \ol s is to study their topological complexity. The first task is to locate $\om$-languages with regard to the Borel and the 
projective hierarchies, and next to the Wadge hierarchy which is a great refinement of the Borel hierarchy.  It is then natural to ask for decidability properties 
and to study decision problems like : is there an effective procedure to determine the Borel rank or the Wadge degree of any context-free $\om$-language ?
Such questions were asked by Lescow and Thomas in \cite{LescowThomas}. In this paper we survey some recent results on the topological complexity of context-free $\om$-languages. 
Some of them were very surprising as the two following ones: 
\begin{enumerate}
\ite there is a $1$-counter finitary language $L$ such that $L^\om$ is analytic but not Borel, \cite{Fin03a}. 
\ite 
The Wadge hierarchy, hence also the Borel hierarchy, 
of $\om$-languages accepted by real time $1$-counter B\"uchi automata is the same as the Wadge hierarchy of $\om$-languages 
accepted by B\"uchi Turing machines, \cite{Fin-mscs06}. 
\end{enumerate}

\noi The Borel and Wadge  hierarchies of {\it non deterministic} context-free $\om$-languages are not effective. One can neither decide whether a given 
context-free $\om$-language is a Borel set nor whether it is in a given Borel class ${\bf \Si}^0_\alpha $ or ${\bf \Pi}^0_\alpha $. 
On the other hand {\it deterministic} context-free $\om$-languages are located at a low level of the Borel hierarchy: they are all 
${\bf \Delta}^0_3$-sets. They  enjoy some  decidability properties although some important questions in this area are still open. We consider 
also the links with the  notions of ambiguity and of  degrees of ambiguity, and the special case of $\om$-powers, i.e. of $\om$-languages in the form 
$V^\om$, where $V$ is a (context-free) finitary language. Finally we state some perspectives and give a list of some  questions which remain open for further 
study. 

\hs The paper is organized as follows. In Section 2 we recall the notions of context-free $\om$-languages accepted by B\"uchi or Muller pushdown automata. 
Topological notions and   Borel and Wadge hierarchies are recalled  in Section 3. 
In Section 4 is studied the case of non-deterministic context-free $\om$-languages while deterministic  context-free $\om$-languages are considered in 
Section 5. Links with notions of ambiguity in context free languages are studied in Section 6.  Section 7 is devoted to the special case of $\om$-powers. 
Perspectives and some open questions are presented in last Section 8.

\section{Context-free $\om$-languages}

We assume the reader to be familiar with the theory of formal ($\om$)-languages  
\cite{Thomas90,Staiger97}.
We shall use usual notations of formal language theory. 
\nl  When $X$ is a finite alphabet, a {\it non-empty finite word} over $X$ is any 
sequence $x=a_1\ldots a_k$, where $a_i\in X$ 
for $i=1,\ldots ,k$ , and  $k$ is an integer $\geq 1$. The {\it length}
 of $x$ is $k$, denoted by $|x|$.
 The {\it empty word} has no letters and is denoted by $\lambda$; its length is $0$. 
 For $x=a_1\ldots a_k$, we write $x(i)=a_i$  
and $x[i]=x(1)\ldots x(i)$ for $i\leq k$ and $x[0]=\lambda$.
 $X^\star$  is the {\it set of finite words} (including the empty word) over $X$.
\nl For $V\subseteq X^\star$, the complement of $V$ (in $X^\star$) is $X^\star - V$ denoted $V^-$.
 \nl  The {\it first infinite ordinal} is $\om$.
 An $\om$-{\it word} over $X$ is an $\om$ -sequence $a_1 \ldots a_n \ldots$, where for all 
integers $ i\geq 1$, ~
$a_i \in X$.  When $\sigma$ is an $\om$-word over $X$, we write
 $\sigma =\sigma(1)\sigma(2)\ldots \sigma(n) \ldots $,  where for all $i$,~ $\sigma(i)\in X$,
and $\sigma[n]=\sigma(1)\sigma(2)\ldots \sigma(n)$  for all $n\geq 1$ and $\sigma[0]=\lambda$.
\nl 
 The usual concatenation product of two finite words $u$ and $v$ is 
denoted $u.v$ (and sometimes just $uv$). This product is extended to the product of a 
finite word $u$ and an $\om$-word $v$: the infinite word $u.v$ is then the $\om$-word such that:
\nl $(u.v)(k)=u(k)$  if $k\leq |u|$, ~~ and ~~ $(u.v)(k)=v(k-|u|)$  if $k>|u|$.
\nl   The {\it prefix relation} is denoted $\sqsubseteq$: a finite word $u$ is a {\it prefix} 
of a finite word $v$ (respectively,  an infinite word $v$), denoted $u\sqsubseteq v$,  
 if and only if there exists a finite word $w$ 
(respectively,  an infinite word $w$), such that $v=u.w$.
 The {\it set of } $\om$-{\it words} over  the alphabet $X$ is denoted by $X^\om$.
An  $\om$-{\it language} over an alphabet $X$ is a subset of  $X^\om$.  The complement (in $X^\om$) of an 
$\om$-language $V \subseteq X^\om$ is $X^\om - V$, denoted $V^-$.
\nl
For $V\subseteq X^\star$, the $\om$-power of $V$ is : 
$$V^\om = \{ \sigma =u_1  \ldots u_n \ldots 
 \in X^\om \mid  \fa i\geq 1~~  u_i\in V \}.$$

\noi We now define  pushdown machines and the class of  $\om$-context-free  \la s.

\begin{Deff}
A pushdown machine (PDM) is a 6-tuple $M=(K,X,\Ga, \delta, q_0, Z_0)$, where $K$ 
is a finite set of states, $X$ is a finite input alphabet, $\Gamma$ is a 
finite pushdown alphabet,
 $q_0\in K$ is the initial state, $Z_0 \in\Ga$ is the start symbol, 
and $\delta$ is a mapping from $K \times (X\cup\{\lambda\} )\times \Ga $ to finite subsets of
$K\times \Gas$ . 
\nl
If  $\gamma\in\Ga^{+}$ describes the pushdown store content, 
the leftmost symbol will be assumed to be on ``top" of the store.
A configuration of a PDM is a pair $(q, \gamma)$ where $q\in K$ and  $\gamma\in\Gas$.\nl
For $a\in X\cup\{\lambda\}$, $\beta,\gamma \in\Ga^{\star}$
and $Z\in\Ga$, if $(p,\beta)$ is in $\delta(q,a,Z)$, then we write
$a: (q,Z\gamma)\mapsto_M (p,\beta\gamma)$.\nl
$\mapsto_M^\star$ is the transitive and reflexive closure of $\mapsto_M$.
(The subscript $M$ will be omitted whenever the meaning remains clear).
\nl
Let $\sigma =a_1a_2 \ldots a_n \ldots $ be an  $\om$-word over $X$. 
An infinite sequence of configurations $r=(q_i,\gamma_i)_{i\geq1}$ is called 
a complete run of $M$ on $\sigma$, starting in configuration $(p,\gamma)$, iff:
\begin{enumerate}
\ite $(q_1,\gamma_1)=(p,\gamma)$

\ite  for each $i\geq 1$, there exists $b_i \in X \cup \{\lambda\}$ 
satisfying $b_i: (q_i,\gamma_i)\mapsto_M(q_{i+1},\gamma_{i+1} )$
such that $a_1a_2 \ldots a_n  \ldots  =b_1b_2 \ldots b_n \ldots $
\end{enumerate}
\noi 
 For every such run, $In(r)$ is the set of all states entered infinitely
 often during run $r$.
\nl
A complete run $r$ of $M$ on $\sigma$ , starting in configuration $(q_0,Z_0)$,
 will be simply called ``a run of $M$ on $\sigma$".
\end{Deff}

\begin{Deff} A B\"uchi pushdown automaton is a 7-tuple
 $M=(K,X,\Ga, \delta, q_0, Z_0, F)$ where $ M'=(K,X,\Ga, \delta, q_0, Z_0)$
is a PDM and $F\subseteq K$ is the set of final states.
The \ol~ accepted by $M$ is 
$$L(M)= \{  \sigma\in X^\om \mid \mbox{ there exists a complete run } r
 \mbox{  of } M  \mbox{ on } \sigma  \mbox{ such that } In(r) \cap F \neq\emptyset \}$$
\end{Deff}

\begin{Deff} A Muller pushdown automaton  is a 7-tuple
 $M=(K,X,\Gamma, \delta, q_0, Z_0, \mathcal{F})$ where $ M'=(K,X,\Gamma, \delta, q_0, Z_0)$
is a PDM and $\mathcal{F}\subseteq 2^K$ is the collection of designated state sets.
The \ol~ accepted by $M$ is 
$$L(M)= \{  \sigma \in X^\om \mid \mbox{ there exists a complete run } r
 \mbox{  of } M  \mbox{ on } \sigma  \mbox{ such that } In(r) \in  \mathcal{F} \}$$
\end{Deff}

\begin{Rem} We consider here two acceptance conditions for $\om$-words, 
the B\"uchi  and the Muller acceptance conditions, respectively denoted 2-acceptance 
and 3-acceptance in \cite{Landweber69} and in \cite{CG78} and $(inf, \sqcap)$ and $(inf, =)$ 
in \cite{Staiger97}.  We refer the reader to \cite{CG,CG78,Staiger97,eh} for consideration of  weaker acceptance conditions, 
and to \cite{2001automata,PerrinPin} for the 
definitions of other usual ones like Rabin, Street, or parity  acceptance conditions. 
Notice however that it seems that the latter ones have not been much considered 
in the study of context-free $\om$-languages but they are often involved in constructions concerning finite automata reading infinite words.  

\end{Rem}

\noi {\bf Notation.}  In the sequel we shall often abbreviate  ``Muller pushdown automaton" by MPDA and ``B\"uchi pushdown automaton" by BPDA. 

\hs Cohen and  Gold and independently Linna established a characterization 
theorem for $\om$-languages accepted by B\"uchi or Muller pushdown automata. 
We shall need the notion of   ``$\om$-Kleene closure"  which  we now firstly  define:

\begin{Deff}
For any family $\mathcal{L}$ of  finitary \la s, the $\om$-Kleene closure
of  $\mathcal{L}$  is : $$\om{\rm-}KC(\mathcal{L}) = \{ \cup_{i=1}^n U_i.V_i^\om  \mid  \fa i\in [1, n] ~~ U_i, V_i \in \mathcal{L}  \}$$
\end{Deff}

\begin{The}[Linna \cite{Linna75}, Cohen and Gold \cite{CG}] \label{theokccf}
Let $CFL$ be the class of context-free (finitary) languages. Then for 
any  \ol~ $L$ the following
three conditions are equivalent:
\begin{enumerate}
\ite $L\in \om{\rm-}KC(CFL)$.
\ite There exists a $BPDA$ that accepts $L$.
\ite There exists a $MPDA$ that accepts $L$.
\end{enumerate}
\end{The}

\noi
In \cite{CG}  are also studied  \ol s generated by $\om$-context-free grammars  
and it is shown that each of the conditions 1), 2), and 3) of the above Theorem is 
also equivalent to: 4) $L$ is generated by a context-free grammar $G$ by leftmost derivations.
These grammars are also studied by Nivat in \cite{Nivat77,Nivat78}.
Then we can let the following definition:

\begin{Deff}
An \ol~ is a context-free $\om$-language   iff it satisfies 
one of the conditions of the above Theorem.
 The class of  context-free $\om$-languages will
 be denoted by $CFL_\om$.
\end{Deff}

\noi If we omit the pushdown store in the above Theorem we  obtain 
the  characterization of  \la s accepted by classical Muller automata (MA) or B\"uchi automata (BA) :

\begin{The}
  For any \ol~ $L$, the following conditions are equivalent:
\begin{enumerate}
\ite   $L$ belongs to $\om{\rm-}KC(REG)$, 
\nl where $REG$ is the class of finitary
 regular languages.
\ite   There exists a MA  that accepts $L$.
\ite     There exists a BA  that accepts $L$.
\end{enumerate}

\noi An \ol~ $L$ satisfying one of the conditions of the above Theorem is called 
 a regular $\om$-language. The class of \orl s will
 be denoted by $REG_\om$.
\end{The}

\noi It follows from   Mc Naughton's Theorem that   the expressive
 power of \de MA (DMA) is equal to the expressive power of non \de MA, i.e. that  
  every regular $\om$-language is accepted by a \de Muller automaton, \cite{Naughton66,PerrinPin}. 
Notice that Choueka gave a simplified proof of Mc Naughton's Theorem in \cite{Choueka74}. Another variant was given by 
Rabin in \cite{Rabin69}. 
\noi Unlike the case of finite automata, \de $MPDA$ do not define the same class
of \ol s as non \de $MPDA$. Let us now define  \de pushdown machines.

\begin{Deff}
A $PDM$ \quad  $M=(K,X,\Ga, \delta, q_0, Z_0)$ is said to be \de  iff for
each $q\in K, Z\in \Ga$, and $a\in X$:
\begin{enumerate}
\ite $\delta(q, a, Z)$ contains at most one element,
\ite $\delta(q, \lambda, Z)$ contains at most one element, and 
\ite if $\delta(q, \lambda, Z)$ is non empty, then $\delta(q, a, Z)$ is empty for all $a\in X$.
\end{enumerate}
\end{Deff}

\noi It turned out that the class of \ol s accepted by \de $BPDA$ is strictly included
into the class of \ol s accepted by \de $MPDA$.  This latest
class is  the class  $DCFL_\om$ of  \de context-free $\omega$-languages. We  denote     $DCFL$ the class 
of \de context-free (finitary) languages.

\begin{Pro}[\cite{CG78}]

\noi
\begin{enumerate}
\ite $DCFL_\om$ is closed under complementation, but is neither closed under union, nor under intersection.
\ite  $DCFL_\om \subsetneq \om{\rm-}KC(DCFL) \subsetneq CFL_\om $ ~~~~
 (these inclusions are strict).
\end{enumerate}
\end{Pro}

\section{Topology}

\subsection{Borel hierarchy and analytic sets}

\noi We assume the reader to be familiar with basic notions of topology which
may be found in \cite{Moschovakis80,LescowThomas,Kechris94,Staiger97,PerrinPin}.
There is a natural metric on the set $X^\om$ of  infinite words 
over a finite alphabet 
$X$ containing at least two letters which is called the {\it prefix metric} and defined as follows. For $u, v \in X^\om$ and 
$u\neq v$ let $\delta(u, v)=2^{-l_{\mathrm{pref}(u,v)}}$ where $l_{\mathrm{pref}(u,v)}$ 
 is the first integer $n$
such that $u(n+1)$  is different from  $v(n+1)$. 
This metric induces on $X^\om$ the usual  Cantor topology for which {\it open subsets} of 
$X^\om$ are in the form $W.X^\om$, where $W\subseteq X^\star$.
A set $L\subseteq X^\om$ is a {\it closed set} iff its complement $X^\om - L$ 
is an open set.
Define now the {\it Borel Hierarchy} of subsets of $X^\om$:

\begin{Deff}
For a non-null countable ordinal $\alpha$, the classes ${\bf \Si}^0_\alpha$
 and ${\bf \Pi}^0_\alpha$ of the Borel Hierarchy on the topological space $X^\om$ 
are defined as follows:
\nl ${\bf \Si}^0_1$ is the class of open subsets of $X^\om$, 
 ${\bf \Pi}^0_1$ is the class of closed subsets of $X^\om$, 
\nl and for any countable ordinal $\alpha \geq 2$: 
\nl ${\bf \Si}^0_\alpha$ is the class of countable unions of subsets of $X^\om$ in 
$\bigcup_{\gamma <\alpha}{\bf \Pi}^0_\gamma$.
 \nl ${\bf \Pi}^0_\alpha$ is the class of countable intersections of subsets of $X^\om$ in 
$\bigcup_{\gamma <\alpha}{\bf \Si}^0_\gamma$.
\end{Deff}

\noi Recall some basic results about these classes :

\begin{Pro}
\noi  
\begin{enumerate}
\ite[(a)] ${\bf \Si}^0_\alpha \cup {\bf \Pi}^0_\alpha  \subsetneq  
{\bf \Si}^0_{\alpha +1}\cap {\bf \Pi}^0_{\alpha +1} $, for each countable 
ordinal  $\alpha \geq 1$. 
\ite[(b)] $\cup_{\gamma <\alpha}{\bf \Si}^0_\gamma = \cup_{\gamma <\alpha}{\bf \Pi}^0_\gamma 
\subsetneq {\bf \Si}^0_\alpha \cap {\bf \Pi}^0_\alpha $, for each countable limit ordinal 
$\alpha$. 
\ite[(c)] A set $W\subseteq X^\om$ is in the class ${\bf \Si}^0_\alpha $ iff its 
complement is in the class ${\bf \Pi}^0_\alpha $. 
\ite[(d)] ${\bf \Si}^0_\alpha  - {\bf \Pi}^0_\alpha  \neq \emptyset $ and 
${\bf \Pi}^0_\alpha  - {\bf \Si}^0_\alpha  \neq \emptyset $ hold 
 for every countable  ordinal $\alpha\geq 1$. 
\end{enumerate}
\end{Pro}

\noi For 
a countable ordinal $\alpha$,  a subset of $X^\om$ is a Borel set of {\it rank} $\alpha$ iff 
it is in ${\bf \Si}^0_{\alpha}\cup {\bf \Pi}^0_{\alpha}$ but not in 
$\bigcup_{\gamma <\alpha}({\bf \Si}^0_\gamma \cup {\bf \Pi}^0_\gamma)$.

\hs    
There are also some subsets of $X^\om$ which are not Borel. 
Indeed there exists another hierarchy beyond the Borel hierarchy, which is called the 
projective hierarchy and which is obtained from  the Borel hierarchy by 
successive applications of operations of projection and complementation.
The first level of the projective hierarchy is formed by the class of {\it analytic sets} and the class of {\it co-analytic sets} which are complements of 
analytic sets.  
In particular 
the class of Borel subsets of $X^\om$ is strictly included into 
the class  ${\bf \Si}^1_1$ of {\it analytic sets} which are 
obtained by projection of Borel sets. 

\begin{Deff} 
A subset $A$ of  $X^\om$ is in the class ${\bf \Si}^1_1$ of {\bf analytic} sets
iff there exist a finite alphabet $Y$ and a Borel subset $B$  of  $(X \times Y)^\om$ 
such that $ x \in A \lra \exists y \in Y^\om $ such that $(x, y) \in B$, 
where $(x, y)$ is the infinite word over the alphabet $X \times Y$ such that
$(x, y)(i)=(x(i),y(i))$ for each  integer $i\geq 1$.
\end{Deff}

\begin{Rem}
In the above definition we could take $B$ in the class ${\bf \Pi}^0_2$. Moreover 
analytic subsets of $X^\om$ are the projections of  ${\bf \Pi}^0_1$-subsets of 
$X^\om \times \om^\om$, where $ \om^\om$ is the Baire space, \cite{Moschovakis80}. 
\end{Rem}

\noi  We now define completeness with regard to reduction by continuous functions. 
For a countable ordinal  $\alpha\geq 1$, a set $F\subseteq X^\om$ is said to be 
a ${\bf \Si}^0_\alpha$  
(respectively,  ${\bf \Pi}^0_\alpha$, ${\bf \Si}^1_1$)-{\it complete set} 
iff for any set $E\subseteq Y^\om$  (with $Y$ a finite alphabet): 
 $E\in {\bf \Si}^0_\alpha$ (respectively,  $E\in {\bf \Pi}^0_\alpha$,  $E\in {\bf \Si}^1_1$) 
iff there exists a continuous function $f: Y^\om \ra X^\om$ such that $E = f^{-1}(F)$. 
 ${\bf \Si}^0_n$
 (respectively ${\bf \Pi}^0_n$)-complete sets, with $n$ an integer $\geq 1$, 
 are thoroughly characterized in \cite{Staiger86a}.  

\hs In particular  $\mathcal{R}=(0^\star.1)^\om$  
is a well known example of 
${\bf \Pi}^0_2 $-complete subset of $\{0, 1\}^\om$. It is the set of 
$\om$-words over $\{0, 1\}$ having infinitely many occurrences of the letter $1$. 
Its  complement 
$\{0, 1\}^\om - (0^\star.1)^\om$ is a 
${\bf \Si}^0_2 $-complete subset of $\{0, 1\}^\om$.

\hs We recall now the definition of the  arithmetical hierarchy of  \ol s which form the effective analogue to the 
hierarchy of Borel sets of finite rank. 
\nl Let $X$ be a finite alphabet. An \ol~ $L\subseteq X^\om$  belongs to the class 
$\Si_n$ if and only if there exists a recursive relation 
$R_L\subseteq (\mathbb{N})^{n-1}\times X^\star$  such that
$$L = \{\sigma \in X^\om \mid Q_1 a_1 Q_2 a_2 \ldots Q_na_n  \quad (a_1,\ldots , a_{n-1}, 
\sigma[a_n+1])\in R_L \}$$

\noi where $Q_1$ is the existential quantifier $\exists$, and every  other $Q_i$, for $2\leq i\leq n$,  is one of the quantifiers $\fa$ or $\exists$ 
(not necessarily in an alternating order). An \ol~ $L\subseteq X^\om$  belongs to the class 
$\Pi_n$ if and only if its complement $X^\om - L$  belongs to the class 
$\Si_n$.  The inclusion relations that hold  between the classes $\Si_n$ and $\Pi_n$ are 
the same as for the corresponding classes of the Borel hierarchy. 
 The classes $\Si_n$ and $\Pi_n$ are  included in the respective classes 
${\bf \Si_n^0}$ and ${\bf \Si_n^0}$ of the Borel hierarchy, and cardinality arguments suffice to show that these inclusions are strict. 

\hs  As in the case of the Borel hierarchy, projections of arithmetical sets 
(of the second $\Pi$-class) lead 
beyond the arithmetical hierarchy, to the analytical hierarchy of \ol s. The first class 
of this hierarchy is the  class $\Si^1_1$ of {\it effective analytic sets} 
 which are obtained by projection of arithmetical sets.
An \ol~ $L\subseteq X^\om$  belongs to the class 
$\Si_1^1$ if and only if there exists a recursive relation 
$R_L\subseteq \mathbb{N} \times \{0, 1\}^\star \times X^\star$  such that:

$$L = \{\sigma \in X^\om  \mid \exists \tau (\tau\in \{0, 1\}^\om \wedge \fa n \exists m 
 ( (n, \tau[m], \sigma[m]) \in R_L )) \}$$

\noi Then an \ol~ $L\subseteq X^\om$  is in the class $\Si_1^1$ iff it is the projection 
of an \ol~ over the alphabet $X\times \{0, 1\}$ which is in the class $\Pi_2$.  The   class $\Pi_1^1$ of  {\it effective co-analytic sets} 
 is simply the class of complements of effective analytic sets. We denote as usual $\Delta_1^1 = \Si^1_1 \cap \Pi_1^1$. 
\nl Recall that an \ol~ $L\subseteq X^\om$ is in the class $\Si_1^1$
iff it is accepted by a non deterministic Turing machine (reading $\om$-words)
with a   B\"uchi or Muller acceptance condition  \cite{Staiger97}. 

\hs  The Borel ranks of  $\Delta_1^1$ sets   are the (recursive) 
ordinals  $\gamma < \om_1^{\mathrm{CK}}$, where $ \om_1^{\mathrm{CK}}$
 is the first non-recursive ordinal, usually called the Church-Kleene ordinal.  
Moreover, for every non null  ordinal $\alpha < \om_1^{\mathrm{CK}}$, there exist some  
${\bf \Si}^0_\alpha$-complete and some  ${\bf \Pi}^0_\alpha$-complete sets in the class $\Delta_1^1$. 
\nl On the other hand,  Kechris, Marker and Sami proved in \cite{KMS89} that the supremum 
of the set of Borel ranks of   (effective)  $\Si_1^1$-sets is the ordinal $\gamma_2^1$. 
 This ordinal is proved to be  strictly greater than the ordinal $\delta_2^1$ which is the first non $\Delta_2^1$ ordinal. 
In particular,   the ordinal            $\gamma_2^1$   is   strictly greater than the ordinal              $ \om_1^{\mathrm{CK}}$.  
Remark that the exact value of the ordinal $\gamma_2^1$ may depend on axioms of  set theory,  see   \cite{KMS89,Fin-mscs06} for more details. 
Notice also that   it seems still unknown  whether {\it every } non null ordinal $\gamma < \gamma_2^1$ is the Borel rank 
of a   $\Si_1^1$-set.

\subsection{Wadge hierarchy}

\noi We now introduce the Wadge hierarchy, which is a great refinement of the Borel hierarchy defined 
via reductions by continuous functions, \cite{Duparc01,Wadge83}. 

\begin{Deff}[Wadge \cite{Wadge83}] Let $X$, $Y$ be two finite alphabets. 
For $L\subseteq X^\om$ and $L'\subseteq Y^\om$, $L$ is said to be Wadge reducible to $L'$
($L\leq _W L')$ iff there exists a continuous function $f: X^\om \ra Y^\om$, such that
$L=f^{-1}(L')$.
\nl $L$ and $L'$ are Wadge equivalent iff $L\leq _W L'$ and $L'\leq _W L$. 
This will be denoted by $L\equiv_W L'$. And we shall say that 
$L<_W L'$ iff $L\leq _W L'$ but not $L'\leq _W L$.
\nl  A set $L\subseteq X^\om$ is said to be self dual iff  $L\equiv_W L^-$, and otherwise 
it is said to be non self dual.
\end{Deff}

\noi
 The relation $\leq _W $  is reflexive and transitive,
 and $\equiv_W $ is an equivalence relation.
\nl The {\it equivalence classes} of $\equiv_W $ are called {\it Wadge degrees}. 
\nl The Wadge hierarchy $WH$ is the class of Borel subsets of a set  $X^\om$, where  $X$ is a finite set,
 equipped with $\leq _W $ and with $\equiv_W $.
\nl  For $L\subseteq X^\om$ and $L'\subseteq Y^\om$, if   
$L\leq _W L'$ and $L=f^{-1}(L')$  where $f$ is a continuous 
function from $ X^\om$  into $Y^\om$, then $f$ is called a continuous reduction of $L$ to 
$L'$. Intuitively it means that $L$ is less complicated than $L'$ because 
to check whether $x\in L$ it suffices to check whether $f(x)\in L'$ where $f$ 
is a continuous function. Hence the Wadge degree of an \ol~
is a measure 
of its topological complexity. 

\hs 
Notice  that in the above definition, we consider that a subset $L\subseteq  X^\om$ is given
together with the alphabet $X$. This is important as it is shown by the following simple example. 
Let $L_1=\{0, 1\}^\om \subseteq \{0, 1\}^\om$ and $L_2=\{0, 1\}^\om \subseteq \{0, 1, 2\}^\om$. So the languages $L_1$ and $L_2$ are equal 
but considered over the different alphabets  $X_1=\{0, 1\}$ and $X_2=\{0, 1, 2\}$. It turns out that $L_1 <_W L_2$. In fact $L_1$ is open {\it and } 
closed in $X_1^\om$ while $L_2$ is closed but non open in $X_2^\om$. 

\hs  We can now define the {\it Wadge class} of a set $L$:

\begin{Deff}
Let $L$ be a subset of $X^\om$. The Wadge class of $L$ is :
$$[L]= \{ L' \mid  L'\subseteq Y^\om \mbox{ for a finite alphabet }Y   \mbox{  and  } L'\leq _W L \}.$$ 
\end{Deff}

\noi Recall that each {\it Borel class} ${\bf \Si}^0_\alpha$ and ${\bf \Pi}^0_\alpha$ 
is a {\it Wadge class}. 
\nl A set $L\subseteq X^\om$ is a ${\bf \Si}^0_\alpha$
 (respectively ${\bf \Pi}^0_\alpha$)-{\it complete set} iff for any set 
$L'\subseteq Y^\om$, $L'$ is in 
${\bf \Si}^0_\alpha$ (respectively ${\bf \Pi}^0_\alpha$) iff $L'\leq _W L $ .
  It follows from the study of the Wadge hierarchy that a set $L\subseteq X^\om$ is a ${\bf \Si}^0_\alpha$
 (respectively, ${\bf \Pi}^0_\alpha$)-{\it complete set} iff it is in ${\bf \Si}^0_\alpha$ but not in ${\bf \Pi}^0_\alpha$
 (respectively, in ${\bf \Pi}^0_\alpha$ but not in ${\bf \Si}^0_\alpha$).

\hs  There is a close relationship between Wadge reducibility
 and games which we now introduce.  

\begin{Deff} Let 
$L\subseteq X^\om$ and $L'\subseteq Y^\om$. 
The Wadge game  $W(L, L')$ is a game with perfect information between two players,
player 1 who is in charge of $L$ and player 2 who is in charge of $L'$.
\nl Player 1 first writes a letter $a_1\in X$, then player 2 writes a letter
$b_1\in Y$, then player 1 writes a letter $a_2\in  X$, and so on. 
\nl The two players alternatively write letters $a_n$ of $X$ for player 1 and $b_n$ of $Y$
for player 2.
\nl After $\om$ steps,  player 1 has written an $\om$-word $a\in X^\om$ and  player 2
has written an $\om$-word $b\in Y^\om$.
 Player 2 is allowed to skip, even infinitely often, provided he really writes an
$\om$-word in  $\om$ steps.
\nl Player 2 wins the play iff [$a\in L \lra b\in L'$], i.e. iff : 
\begin{center}
  [($a\in L ~{\rm and} ~ b\in L'$)~ {\rm or} ~ 
($a\notin L ~{\rm and}~ b\notin L'~{\rm and} ~ b~{\rm is~infinite}  $)].
\end{center}
\end{Deff}

\noi
Recall that a strategy for player 1 is a function 
$\sigma :(Y\cup \{s\})^\star\ra X$.
And a strategy for player 2 is a function $f:X^+\ra Y\cup\{ s\}$.
\nl $\sigma$ is a winning stategy  for player 1 iff he always wins a play when
 he uses the strategy $\sigma$, i.e. when the  $n^{th}$  letter he writes is given
by $a_n=\sigma (b_1\ldots b_{n-1})$, where $b_i$ is the letter written by player 2 
at step $i$ and $b_i=s$ if player 2 skips at step $i$.
\nl A winning strategy for player 2 is defined in a similar manner.

\hs   Martin's Theorem states that every Gale-Stewart Game $G(B)$,  with $B$ a Borel set, 
is determined, i.e. that one of  the two players has a winning strategy in the game $G(B)$, see \cite{Kechris94}. 
This implies the following determinacy result :

\begin{The} [Wadge] Let $L\subseteq X^\om$ and $L'\subseteq Y^\om$ be two Borel sets, where
$X$ and $Y$ are finite  alphabets. Then the Wadge game $W(L, L')$ is determined :
one of the two players has a winning strategy. And $L\leq_W L'$ iff  player 2 has a 
winning strategy  in the game $W(L, L')$.
\end{The}

\begin{The} [Wadge]\label{wh}
Up to the complement and $\equiv _W$, the class of Borel subsets of $X^\om$,
 for  a finite alphabet $X$, is a well ordered hierarchy.
 There is an ordinal $|WH|$, called the length of the hierarchy, and a map
$d_W^0$ from $WH$ onto $|WH|-\{0\}$, such that for all $L, L' \subseteq X^\om$:
\nl $d_W^0 L < d_W^0 L' \lra L<_W L' $  and 
\nl $d_W^0 L = d_W^0 L' \lra [ L\equiv_W L' $ or $L\equiv_W L'^-]$.
\end{The}

\noi 
 The Wadge hierarchy of Borel sets of {\it finite rank }
has  length $^1\varepsilon_0$ where $^1\varepsilon_0$
 is the limit of the ordinals $\alpha_n$ defined by $\alpha_1=\om_1$ and 
$\alpha_{n+1}=\om_1^{\alpha_n}$ for $n$ a non negative integer, $\om_1$
 being the first non countable ordinal. Then $^1\varepsilon_0$ is the first fixed 
point of the ordinal exponentiation of base $\om_1$. The length of the Wadge hierarchy 
of Borel sets in ${\bf \Delta^0_\om}= {\bf \Si^0_\om}\cap {\bf \Pi^0_\om}$ 
  is the $\om_1^{th}$ fixed point 
of the ordinal exponentiation of base $\om_1$, which is a much larger ordinal. The length 
of the whole Wadge hierarchy of Borel sets is a huge ordinal, with regard 
to the $\om_1^{th}$ fixed point 
of the ordinal exponentiation of base $\om_1$. It is described in \cite{Wadge83,Duparc01} 
by the use of the Veblen functions.

\section{Topological complexity of context-free $\om$-languages}\label{cf}

We recall first results about the  topological complexity of regular $\om$-languages. 
Topological properties of \orl s were first
 studied by L. H. Landweber in \cite{Landweber69} where he characterized  
 \orl s  in a given Borel class.
It turned out that a regular $\om$-language is a ${\bf \Pi}^0_2$-set iff it is accepted by a \de B\"uchi automaton. 
On the other hand 
Mc Naughton's Theorem implies that \orl s,  accepted by \de Muller automata,  are boolean 
combinations of regular $\om$-languages accepted by \de B\"uchi automata. Thus they are boolean combinations of ${\bf \Pi}^0_2$-sets hence 
${\bf \Delta}^0_3$-sets. Moreover Landweber proved that one can effectively determine the exact level of a given regular $\om$-language
with regard to the Borel hierarchy. 
\nl A great improvement of these results was obtained by Wagner who determined in an effective way,  using the notions of chains and superchains, 
 the Wadge hierarchy of the class $REG_\om$, \cite{Wagner79}. 
This hierarchy has length $\om^\om$ and is now called the Wagner hierarchy, \cite{Selivanov95,Selivanov03b,Selivanov03a,Selivanov98,Staiger97}. 
Wilke and Yoo  
proved in \cite{WilkeYoo95} that  one can compute in polynomial time the Wadge degree of a \orl. 
Later Carton and Perrin gave a presentation of the Wagner hierarchy using algebraic notions of $\om$-semigroups,  
\cite{CartonPerrin99,CartonPerrin97b,PerrinPin}. This work was 
completed by Duparc and Riss in \cite{DR}. 

\hs  Context-free $\om$-languages beyond the class ${\bf \Delta}^0_3$ have been constructed for the first time in \cite{Fin01a}. 
The construction used an operation of exponentiation of sets of finite or 
infinite words introduced by Duparc in his study of the Wadge hierarchy \cite{Duparc01}. We are going now to recall these constructions  although 
some stronger results on the topological complexity of context-free $\om$-languages were obtained later in \cite{cie05,Fin-mscs06} by other methods. However 
the methods of \cite{Fin01a} using Duparc's operation of exponentiation are also interesting and it gave other results on ambiguity and 
on $\om$-powers of context-free languages we can not (yet ?) get by other methods, see Sections \ref{section-amb} and \ref{powers} below. 

\hs 
Wadge gave   a description of the Wadge hierarchy of Borel sets in \cite{Wadge83}.
Duparc recently got a new proof of Wadge's results and gave in \cite{Duparc-phd,Duparc01}  a normal form of Borel sets in the class ${\bf \Delta}_\om^0$,  
i.e. an inductive construction of a Borel set of every given degree smaller than  the  $\om_1^{th}$ fixed point 
of the ordinal exponentiation of base $\om_1$. The construction 
relies on set theoretic operations which are the counterpart of arithmetical operations over 
ordinals needed to compute the Wadge degrees. 
\nl  Actually  Duparc studied the Wadge hierarchy via the study of the conciliating hierarchy. 
Conciliating sets  are sets of 
finite {\it or} infinite words over an alphabet $X$, i.e. subsets of 
$X^\star \cup X^\om = X^{\leq \om}$.
It turned out that the conciliating  hierarchy is isomorphic to the Wadge hierarchy 
of non-self-dual Borel sets, via the  correspondence $A \ra A^d$ we  recall now: 

\hs For a word $ x\in (X \cup \{d\})^{\leq \om}$  we denote by  $x( /d)$ the sequence obtained from $x$ 
by removing every occurrence of the letter $d$. Then 
for $A\subseteq X^{\leq \om}$ and $d$ a letter not in $X$, $A^d$ is the $\om$-language over  $X \cup \{d\}$  
which is defined by : 
 $$A^d = \{ x\in (X \cup \{d\})^\om \mid x( /d)\in A \}.$$

\noi 
We are going now to introduce the operation of exponentiation of
conciliating sets.

\begin{Deff}[Duparc \cite{Duparc01}]\label{til}
Let  $X$ be a finite alphabet, $\tla  \notin X$,  and let 
 $x$ be a finite or infinite word over the alphabet $Y=X\cup \{\tla\}$.
\nl Then  $x^\tla$ is inductively defined by:
\nl $\lambda^\tla =\lambda$,
\nl and for a finite word $u\in (X\cup \{\tla\})^\star$:
\nl $(u.a)^\tla=u^\tla.a$, if $a\in X$,
\nl  $(u.\tla)^\tla =u^\tla(1).u^\tla(2)\ldots u^\tla(|u^\tla|-1)$  if
$|u^\tla|>0$,
\nl $(u.\tla)^\tla=\lambda$  if $|u^\tla|=0$,
\nl and for $u$ infinite:
\nl $(u)^\tla = \lim_{n\in\om} (u[n])^\tla$, where, given $\beta_n$ and $v$
in   $X^\star$,
\nl $v\sqsubseteq \lim_{n\in\om} \beta_n \lra  \exists n \fa p\geq n\quad
\beta_p[|v|]=v$.
\nl(The finite {\it or} infinite word $\lim_{n\in\om} \beta_n$ is
determined by the set of its (finite) prefixes).
\end{Deff}

\begin{Rem}
For $x \in Y^{\leq \om}$, $x^\tla$ denotes the string $x$, once every $\tla$
occuring in $x$
has been ``evaluated" to the back space operation, 
proceeding from left to right inside $x$. In other words $x^\tla = x$ from
which every
 interval of the form $``a\tla "$ ($a\in X$) is removed.
\end{Rem}

\noi For example if $u=(a\tla)^n$, for $n$ an integer $\geq 1$, or
$u=(a\tla)^\om$,  or $u=(a\tla\tla)^\om$,  then $(u)^\tla=\lambda$.
If $u=(ab\tla)^\om$ then $(u)^\tla=a^\om$  and
 if $u=bb(\tla a)^\om$ then $(u)^\tla=b$.

\hs Let us notice that in Definition \ref{til} the limit is not defined in
the usual way:
\nl for example if $u=bb(\tla a)^\om$ the finite word  $u[n]^\tla$ is
alternatively
equal to $b$ or to $ba$: more precisely $u[2n+1]^\tla=b$ and
$u[2n+2]^\tla=ba$ for every
integer $n\geq 1$ (it holds also that
$u[1]^\tla=b$ and  $u[2]^\tla=bb$). Thus Definition \ref{til} implies that
$\lim_{n\in\om} (u[n])^\tla = b$ so $u^\tla=b$.
\nl  We can now define the operation $A \ra A^\sim$ of
{\it exponentiation of conciliating sets}:

\begin{Deff}[Duparc \cite{Duparc01}]
For $A\subseteq X^{\leq \om}$ and $\tla \notin X$, let 
$$A^\sim =_{df} \{x\in (X\cup \{\tla\})^{\leq \om} \mid  x^\tla\in
A\}.$$
\end{Deff}

\noi The operation $\sim$ is monotone with regard to the Wadge ordering and
produces some sets
of higher complexity.

\begin{The} [Duparc \cite{Duparc01} ] \label{thedup}
Let $A\subseteq X^{\leq \om}$ and  $n\geq 1$. if  $A^d\subseteq (X \cup\{d\})^\om$ is a ${\bf \Si}_n^0$-complete
 (respectively, ${\bf \Pi}_n^0$-complete) set, 
then $(A^\sim)^d$ is a 
${\bf \Si}_{n+1}^0$-complete
 (respectively,  ${\bf \Pi}_{n+1}^0$-complete) set.
\end{The}

\noi It was proved in \cite{Fin01a} that the class of context-free infinitary languages (which are unions of a context-free finitary language 
and of a context-free $\om$-language) is closed under the operation $A \ra A^\sim$. On the other hand  $A\ra A^d$ is an operation from 
the class of context-free infinitary languages into the class of context-free $\om$-languages. 
This implies that, for each integer $n\geq 1$, there exist some context-free $\om$-languages which are ${\bf \Si}_n^0$-complete and some 
others which are ${\bf \Pi}_n^0$-complete. 

\begin{The}[\cite{Fin01a}]\label{the-cf}
For each non negative integer $n\geq 1$, there exist  ${\bf \Si}_n^0$-complete context-free $\om$-languages $A_n$
and ${\bf \Pi}_n^0$-complete context-free $\om$-languages $B_n$.
\end{The}

\noi {\bf Proof.}  For $n=1$  consider the ${\bf \Si}_1^0$-complete \orl~ 
\nl $A_1=\{ \alpha \in\{0, 1\}^\om  \mid  \exists i\quad \alpha (i)=1\}$
\nl and the ${\bf \Pi}_1^0$-complete \orl~  
\nl $B_1=\{ \alpha \in\{0, 1\}^\om    \mid \fa  i\quad \alpha (i)=0\}$.
\nl These \la s are context-free $\om$-languages  because $REG_\om \subseteq CFL_\om$.
\nl  Now consider the ${\bf \Si}_2^0$-complete \orl~ 
\nl $A_2=\{ \alpha \in\{0, 1\}^\om   \mid  \exists^{<\om} i\quad \alpha (i)=1\}$
\nl and the ${\bf \Pi}_2^0$-complete \orl~  
\nl $B_2=\{ \alpha \in\{0, 1\}^\om   \mid  \exists^{\om}  i\quad \alpha (i)=0\}$,
\nl where $\exists^{<\om} i$ means: " there exist only finitely many $i$ such that $\ldots$" , and 
\nl $\exists^{\om} i$ means: " there exist  infinitely many $i$ such that $\ldots$". 
\nl $A_2$ and $B_2$ are  context-free $\om$-languages   because they are \orl s.

\hs To obtain context-free $\om$-languages   of greater  Borel ranks, consider 
 now $O_1$ (respectively,  $C_1$ ) subsets of $\{0, 1\}^{\leq \om}$ such that
 $(O_1)^d$ (respectively,  $(C_1)^d$ ) are   
 ${\bf \Si}_1^0$-complete
( respectively ${\bf \Pi}_1^0$-complete ) .
\nl For example $O_1=\{ x\in \{0, 1\}^{\leq \om}  \mid  \exists ~i ~x(i)=1\}$ and 
  $C_1=\{ \lambda\}$.

\hs  We can apply $n\geq 1 $ times the 
operation of exponentiation of sets.
\nl More precisely, we define, for a set $A\subseteq X^{\leq \om}$:
\nl $A^{\sim .0}=A$
\nl $A^{\sim .1}=A^\sim$  and
\nl $A^{\sim .(n+1)}=(A^{\sim .n})^\sim$ .

\hs Now  apply $n$ times (for an integer $n\geq 1$)  the operation $\sim$ 
(with different new letters 
$\tla_1$, $\tla_2$, $\tla_3$, \ldots , $\tla_{n}$) to  $O_1$ and $C_1$.

\hs By Theorem \ref{thedup}, it holds that for an integer $n\geq 1$:
\nl $(O_1^{\sim .n})^d $ is a ${\bf \Si}_{n+1}^0$-complete 
subset of $\{0, 1, \tla_1,\ldots , \tla_n, d\}^\om$. 
\nl $(C_1^{\sim .n})^d $ is a ${\bf \Pi}_{n+1}^0$-complete 
subset of $\{0, 1, \tla_1,\ldots  , \tla_n, d\}^\om$. 

\hs And  it is easy to see that $O_1$ and $C_1$ are in the form  
$ E \cup F$  where $E$ is a finitary context-free language and $F$ is a  context-free 
$\om$-language. Then  the \ol s  $(O_1^{\sim .n})^d $ and $(C_1^{\sim .n})^d $ are context-free. Hence the 
 class $CFL_\om$  exhausts the finite ranks of the Borel hierarchy:
we  obtain the context-free $\om$-languages   
 $A_n=(O_1^{\sim .(n-1)})^d $ and $B_n=(C_1^{\sim .(n-1)})^d $,
for $n\geq 3$.
\ep

\hs This gave a partial answer to questions of Thomas and Lescow \cite{LescowThomas} about the hierarchy of 
context-free $\om$-languages.
\nl A natural question now arose: Do the decidability results of \cite{Landweber69} extend to context-free $\om$-languages? 
Unfortunately the answer is  no.
 Cohen and Gold proved that one cannot decide whether a given  context-free $\om$-language  is in the class 
${\bf \Pi}_1^0$, ${\bf \Si}_1^0$, or ${\bf \Pi}_2^0$, \cite{CG}.  This result was first extended to all classes ${\bf \Si}_n^0$ and
${\bf \Pi}_n^0$, for $n$ an integer $\geq 1$, using the undecidability of the Post Correspondence Problem, \cite{Fin01a}.

\hs Later,  the  coding of an infinite number of erasers $\tla_n$, $n\geq 1$, and an iteration of the operation of exponentiation were used 
 to prove that there exist some context-free $\om$-languages which are Borel of infinite rank, \cite{Fin03c}.

\hs Using the correspondences between the operation of exponentiation of sets and the ordinal exponentiation of base $\om_1$, and between the Wadge's 
operation of sum of sets,   \cite{Wadge83,Duparc01}, and the ordinal sum, it was proved in \cite{Fin01b} that the length of the Wadge hierarchy of the 
class $CFL_\om$ is at least $\varepsilon_0$, the first fixed point of the ordinal exponentiation of base $\om$. 
Next were constructed some ${\bf \Delta}_\om^0$ context-free 
$\om$-languages in $\varepsilon_\om$ Wadge degrees, 
where $\varepsilon_\om$ is the $\om^{th}$  fixed point of the ordinal exponentiation of base $\om$,   and 
also some ${\bf \Sigma}_\om^0$-complete context-free 
$\om$-languages, 
\cite{Fin01c,Fin05}. Notice that the Wadge hierarchy of {\it non-deterministic}
context-free $\om$-languages  is not effective, \cite{Fin01b}. 

\hs The question of the existence of non-Borel context-free $\om$-languages was solved by Finkel and Ressayre. Using a coding of infinite binary trees 
labeled in a finite alphabet $X$, it was proved that there exist some non-Borel, and even ${\bf \Si}^1_1$-complete, context-free 
$\om$-languages, and that one cannot decide 
whether a given context-free $\om$-language is a Borel set,  \cite{Fin03a}.  Amazingly there is a simple finitary language $V$ accepted by a $1$-counter 
automaton such that $V^\om$ is ${\bf \Si}^1_1$-complete; we shall recall  it in Section \ref{powers} below on $\om$-powers.

\hs But a complete and  very surprising result was obtained in \cite{cie05,Fin-mscs06}, which extended  previous results.  
A simulation of  multicounter automata by  $1$-counter automata was used in \cite{cie05,Fin-mscs06}. We firstly recall now the definition 
of these automata, in order to sketch the constructions involved in these simulations. 

\begin{Deff} Let $k$ be an integer $\geq 1$. 
A  $k$-counter machine ($k$-CM) is a 4-tuple 
$\mathcal{M}$=$(K,X, \Delta, q_0)$,  where $K$ 
is a finite set of states, $X$ is a finite input alphabet, 
 $q_0\in K$ is the initial state, 
and  $\Delta \subseteq K \times ( X \cup \{\lambda\} ) \times \{0, 1\}^k \times K \times \{0, 1, -1\}^k$ is the transition relation. 
The $k$-counter machine $\mathcal{M}$ is said to be {\it real time} iff: 
$\Delta \subseteq K \times
  X \times \{0, 1\}^k \times K \times \{0, 1, -1\}^k$, 
 i.e. iff there are not any $\lambda$-transitions. 
\nl  
If  the machine $\mathcal{M}$ is in state $q$ and 
$c_i \in \mathbf{N}$ is the content of the $i^{th}$ counter 
 $\mathcal{C}$$_i$ then 
the  configuration (or global state)
 of $\mathcal{M}$ is the  $(k+1)$-tuple $(q, c_1, \ldots , c_k)$.

\hs For $a\in X \cup \{\lambda\}$, 
$q, q' \in K$ and $(c_1, \ldots , c_k) \in \mathbf{N}^k$ such 
that $c_j=0$ for $j\in E \subseteq  \{1, \ldots , k\}$ and $c_j >0$ for 
$j\notin E$, if 
$(q, a, i_1, \ldots , i_k, q', j_1, \ldots , j_k) \in \Delta$ where $i_j=0$ for $j\in E$ 
and $i_j=1$ for $j\notin E$, then we write:
$$a: (q, c_1, \ldots , c_k)\mapsto_{\mathcal{M}} (q', c_1+j_1, \ldots , c_k+j_k)$$

\noi Thus we see that the transition relation must satisfy:
 \nl if $(q, a, i_1, \ldots , i_k, q', j_1, \ldots , j_k)  \in    \Delta$ and  $i_m=0$ for 
 some $m\in \{1, \ldots , k\}$, then $j_m=0$ or $j_m=1$ (but $j_m$ cannot be equal to $-1$).

\hs
Let $\sigma =a_1a_2 \ldots a_n \ldots $ be an $\om$-word over $X$. 
An $\om$-sequence of configurations $r=(q_i, c_1^{i}, \ldots c_k^{i})_{i \geq 1}$ is called 
a run of $\mathcal{M}$ on $\sigma$, starting in configuration 
$(p, c_1, \ldots, c_k)$, iff:
\begin{enumerate}
\ite[(1)]  $(q_1, c_1^{1}, \ldots c_k^{1})=(p, c_1, \ldots, c_k)$

\ite[(2)]   for each $i\geq 1$, there  exists $b_i \in X \cup \{\lambda\}$ such that
 $b_i: (q_i, c_1^{i}, \ldots c_k^{i})\mapsto_{\mathcal{M}}  
(q_{i+1},  c_1^{i+1}, \ldots c_k^{i+1})$  
such that either ~  $a_1a_2\ldots a_n\ldots =b_1b_2\ldots b_n\ldots$ 
\nl or ~  $b_1b_2\ldots b_n\ldots$ is a finite prefix of ~ $a_1a_2\ldots a_n\ldots$
\end{enumerate}
\noi The run $r$ is said to be complete when $a_1a_2\ldots a_n\ldots =b_1b_2\ldots b_n\ldots$ 
\nl 
For every such run, $\mathrm{In}(r)$ is the set of all states entered infinitely
 often during run $r$.
\nl
A complete run $r$ of $M$ on $\sigma$, starting in configuration $(q_0, 0, \ldots, 0)$,
 will be simply called ``a run of $M$ on $\sigma$".
\end{Deff}

\begin{Deff} A B\"uchi $k$-counter automaton  is a 5-tuple 
$\mathcal{M}$=$(K,X, \Delta, q_0, F)$, 
where $ \mathcal{M}'$=$(K,X, \Delta, q_0)$
is a $k$-counter machine and $F \subseteq K$ 
is the set of accepting  states.
The \ol~ accepted by $\mathcal{M}$ is 
\begin{center}
$L(\mathcal{M})$= $\{  \sigma\in X^\om \mid \mbox{  there exists a  run r
 of } \mathcal{M} \mbox{ on } \sigma \mbox{  such that } \mathrm{In}(r)
 \cap F \neq \emptyset \}$
\end{center}
\end{Deff}

\noi The notion of  Muller $k$-counter automaton  is defined in a similar way. 
One can see that an  $\om$-language is accepted by a (real time) 
B\"uchi $k$-counter automaton iff it is accepted by a 
(real time) Muller  $k$-counter automaton \cite{eh}. Notice that this result is no longer true in the  deterministic case.    
\nl  We denote ${\bf BC}(k)$ (respectively,  {\bf r}-${\bf BC}(k)$)    the class of  B\"uchi $k$-counter automata  (respectively, 
 of real time B\"uchi $k$-counter automata. 
\nl We denote ${\bf BCL}(k)_\om$    (respectively,   {\bf r}-${\bf BCL}(k)_\om$)   the class of \ol s accepted by  B\"uchi $k$-counter automata  
 (respectively,   by real time B\"uchi $k$-counter automata). 

\hs 
  Remark that  $1$-counter automata   introduced above are equivalent to pushdown automata 
whose stack alphabet is in the form $\{Z_0, A\}$ where $Z_0$ is the bottom symbol which always 
remains at the bottom of the stack and appears only there and $A$ is another stack symbol. 
The pushdown stack may be seen like a counter whose content is the integer $N$ if the stack 
content is the word $A^N.Z_0$. 
\nl In the model introduced here the counter value cannot be increased by  more than 1 during  
a single transition. However this does not change the class of $\om$-languages accepted 
by such automata. So the class ${\bf BCL}(1)_\om$ is equal to the class 
{\bf 1}-${\bf ICL_\om}$, introduced in \cite{Fin01b},  
and it  is a strict subclass of the class ${\bf CFL}_\om$ of context-free \ol s
accepted by B\"uchi pushdown automata.

\hs We state now the surprising  result proved in \cite{Fin-mscs06}, using multicounter-automata.

\begin{The}[\cite{Fin-mscs06}]\label{thewad}    
The Wadge hierarchy of the class 
{\bf r}-${\bf BCL}(1)_\om$, hence also of the class ${\bf CFL}_\om$, or of every 
class $\mathcal{C}$ such that 
$\mbox{ {\bf r}-}{\bf BCL}(1)_\om \subseteq  \mathcal{C} $$ \subseteq  \Sigma^1_1$, 
is the Wadge hierarchy of the class 
 $\Sigma^1_1$ of $\om$-languages accepted by Turing machines with a B\"uchi acceptance 
condition. 
\end{The}

\noi We now sketch the proof of this result.  It is well known that every Turing machine can be simulated by a 
(non real time) $2$-counter automaton, see \cite{HopcroftUllman79}.  
Thus the Wadge hierarchy of the class  ${\bf BCL}(2)_\om$ is also the Wadge  hierarchy of the class 
of $\om$-languages accepted by B\"uchi Turing machines. 
\nl 
One can  then find, from an $\om$-language $L \subseteq X^\om$ in ${\bf BCL}(2)_\om$, another 
$\om$-language $\theta_S(L)$ which will be of the same topological complexity but  accepted by a {\it real-time } 8-counter B\"uchi automaton.  
The idea is to add firstly a storage type called a queue to a 2-counter B\"uchi automaton in order to read $\om$-words in real-time. Then  
the queue can be simulated by two pushdown stacks or by four counters. This simulation is not done in real-time but a crucial fact is that one can bound 
the number of transitions needed to simulate the queue. This allows to pad the strings in $L$ with enough extra letters so that the new words will be read in 
real-time by a 8-counter B\"uchi automaton.   The padding is obtained via the function $\theta_S$ which we define now. 

\hs Let $X$ be an alphabet having at least two letters,  $E$ be a new letter not in 
$X$,  $S$ be an integer $\geq 1$, and $\theta_S: X^\om \ra (X \cup \{E\})^\om$ be the 
function defined, for all  $x \in X^\om$, by: 
$$ \theta_S(x)=x(1).E^{S}.x(2).E^{S^2}.x(3).E^{S^3}.x(4) \ldots 
x(n).E^{S^n}.x(n+1).E^{S^{n+1}} \ldots $$

\hs It turns out that if $L \subseteq X^\om$ is in ${\bf BCL}(2)_\om$ then there exists an integer $S \geq 1$ such 
that $\theta_S(L)$ is in the class  {\bf r}-${\bf BCL}(8)_\om$, and,   except for some  special few cases,    $\theta_S(L) \equiv_W L$. 

\hs The next step is to simulate a {\it real-time } 8-counter B\"uchi automaton, using only a {\it real-time } 1-counter B\"uchi automaton. 

\hs Consider the product of the eight first prime numbers: 
$$K = 2\times 3\times 5\times 7\times 11\times 13\times 17\times 19 = 9699690$$
\noi Then an $\om$-word $x\in X^\om$ can  be coded by the $\om$-word 
$$h(x)=A.0^K.x(1).B.0^{K^2}.A.0^{K^2}.x(2).B.0^{K^3}.A.0^{K^3}.x(3).B \ldots  
B.0^{K^n}.A.0^{K^n}.x(n).B \ldots  $$

\noi where  
$A$, $B$ and $0$  are  new letters not in $X$. The mapping $h : X^\om \ra  (X \cup\{A, B, 0\})^\om$  is continuous.  
It is easy to see that the $\om$-language $h(X^{\om})^-$ is an open subset of     $(X \cup\{A, B, 0\})^\om$  and that  it is    in the class 
{\bf r}-${\bf BCL}(1)_\om$. 

\hs If     $L(\mathcal{A})  \subseteq X^\om$     is accepted by a 
 real time $8$-counter B\"uchi automaton  $\mathcal{A}$, then  one can construct  effectively from $\mathcal{A}$ a 
$1$-counter B\"uchi automaton $\mathcal{B}$, reading words over the alphabet $X \cup\{A, B, 0\}$, 
 such that          $L(\mathcal{A}) $$= h^{-1}( L(\mathcal{B}) )$, i.e. 
$$\fa x \in X^{\om} ~~~~ h(x) \in L(\mathcal{B}) \longleftrightarrow 
x\in  L(\mathcal{A})$$ 

\noi In fact, the  simulation,    during the reading of $h(x)$ by the  
$1$-counter B\"uchi automaton $\mathcal{B}$,  of the behaviour of the real time 
$8$-counter B\"uchi automaton  $\mathcal{A}$ reading  $x$, 
 can be achieved, using   the coding of 
 the content $(c_1, c_2, \ldots, c_8)$ 
of eight counters  by the product $2^{c_1}\times 3^{c_2}\times  \ldots \times (17)^{c_7}\times (19)^{c_8}$, 
 and the {\bf special shape} of $\om$-words 
in $h(X^\om)$ which allows the propagation of the value of the counters  of $\mathcal{A}$. 
A crucial fact here is that  $h(X^\om)^-$ is in the class {\bf r}-${\bf BCL}(1)_\om$. Thus the $\om$-language 
$$    h(  L(\mathcal{A})   )    \cup h(X^{\om})^- =   L(\mathcal{B}) \cup h(X^{\om})^-$$
\noi is in the class ${\bf BCL}(1)_\om$ and it has the same topological complexity as the $\om$-language $L(\mathcal{A}) $, 
(except the special few cases where $d_W(L(\mathcal{A}))\leq \om$).

\hs One can see, from the construction of $\mathcal{B}$, that  at most $(K-1)$ consecutive $\lambda$-transitions  can  occur during the reading 
of an $\om$-word $x$ by  $\mathcal{B}$. It is then easy to see that the 
 $\om$-language $\phi ( h( L(\mathcal{A}) )$$ \cup h(X^{\om})^- )$ is an $\om$-language 
in the class  {\bf r}-${\bf BCL}(1)_\om$ which has the same topological complexity as the $\om$-language $L(\mathcal{A}) $, where 
$\phi$ is   the  mapping from    $(X \cup\{A, B, 0\})^\om$ into  $ (X \cup\{A, B, F,  0\})^\om $, with $F$  a new letter, which  is defined  by:  
$$\phi(x) = F^{K-1}.x(1).F^{K-1}.x(2).F^{K-1}.x(3) 
\ldots F^{K-1}.x(n). F^{K-1}.x(n+1).F^{K-1} \ldots$$

\noi Altogether these constructions are used in  \cite{Fin-mscs06}  to prove Theorem \ref{thewad}.  
As the  Wadge hierarchy is a refinement of the Borel hierarchy and, for any countable ordinal $\alpha$, 
${\bf \Si}^0_\alpha$-complete sets (respectively, ${\bf \Pi}^0_\alpha$-complete sets) 
form a single Wadge degree, this implies also the following result.  

\begin{The}\label{thebor}    

\noi Let $\mathcal{C}$ be a class of $\om$-languages such that:  
\begin{center}
$\mbox{ {\bf r}-}{\bf BCL}(1)_\om \subseteq  \mathcal{C}$$  \subseteq  \Sigma^1_1.$
\end{center}
\begin{enumerate} 
\ite[(a)]  The Borel hierarchy of the class $\mathcal{C}$ is equal to the Borel hierarchy of the class $\Sigma^1_1$. 
\ite[(b)]   $\gamma_2^1=Sup ~~ \{ \alpha \mid \exists L \in \mathcal{C} \mbox{ such that }$$L $$\mbox{ is a Borel set of rank } \alpha \}.$  
\ite[(c)]  For every non null  ordinal $\alpha < \om_1^{\mathrm{CK}}$, 
there exists some  
${\bf \Si}^0_\alpha$-complete and some ${\bf \Pi}^0_\alpha$-complete
$\om$-languages in the class $\mathcal{C}$.  
\end{enumerate}
\end{The}

\noi Notice that similar methods have next be used to get another surprising result: the Wadge hierarchy, hence also the Borel hierarchy, 
of infinitary rational relations accepted by $2$-tape B\"uchi automata is equal to the Wadge hierarchy 
of the class $\mbox{ {\bf r}-}{\bf BCL}(1)_\om$ or of the class $\Sigma^1_1$, \cite{Fin06b,Fink-Wd}.

\section{Topological complexity of deterministic context-free $\om$-languages}

\noi We have seen in the previous section that all {\it non-deterministic} 
finite machines accept $\om$-languages of the same topological complexity, as soon as they 
can simulate a real time $1$-counter automaton. 
\nl  This result is  still true in the {\it deterministic} case if we consider only the Borel hierarchy.  
 Recall that regular $\om$-languages accepted by B\"uchi automata are ${\bf \Pi}^0_2$-sets and $\om$-languages accepted by Muller automata
are boolean combinations of ${\bf \Pi}^0_2$-sets hence 
${\bf \Delta}^0_3$-sets. 
 Engelfriet and  Hoogeboom proved that  this result holds also for all \ol s  accepted by {\it  \de }
 {\bf X}-automata, i.e. automata equipped with a storage type {\bf X}, including the cases of 
$k$-counter automata,  pushdown automata, Petri nets, Turing machines. 
In particular,  $\om$-languages accepted by deterministic B\"uchi   Turing machines   are ${\bf \Pi}^0_2$-sets     and   
$\om$-languages accepted by deterministic   Muller Turing machines are ${\bf \Delta}^0_3$-sets. 

\hs It turned out that this  is  no longer true 
 if we consider  the much finer Wadge hierarchy to measure the complexity of $\om$-languages. The Wadge hierarchy is suitable to distinguish the 
accepting power of deterministic finite machines reading infinite words. 
Recall that the Wadge hierarchy of regular $\om$-languages, now called the Wagner hierarchy, has been 
effectively determined by Wagner; it has length $\om^\om$ \cite{Wagner79,Selivanov95,Selivanov98}. 

\hs Its extension to   {\it deterministic}      context-free $\om$-languages has been determined by Duparc, its
 length is $\om^{(\om^2)}$ \cite{DFR,Duparc03}. To determine the Wadge hierarchy of the class $DCFL_\om$, Duparc first defined operations on 
DMPDA which correspond to ordinal operations of sum, multiplication by $\om$, and multiplication by $\om_1$, over Wadge degrees. In this way are
constructed  some  DMPDA accepting $\om$-languages of every Wadge degree in the form : 

$$d_W^0(A) = \om_1^{n_j}.\delta_j + \om_1^{n_{j-1}}.\delta_{j-1} + \ldots 
 + \om_1^{n_1}.\delta_1$$

\noi where  $j>0$ is an integer,
$n_j > n_{j-1} > \ldots  > n_1 $  are integers $\geq 0$, and 
$\delta_j, \delta_{j-1}, \ldots , \delta_1$ are non null ordinals $<\om^\om$.

\hs On the other hand it is known that the Wadge degree $\alpha$ of a boolean combination of ${\bf \Pi}^0_2$-sets is smaller than the ordinal 
$\om_1^\om$ thus it has a Cantor normal form : 
$$\alpha = \om_1^{n_j}.\delta_j + \om_1^{n_{j-1}}.\delta_{j-1} + \ldots 
 + \om_1^{n_1}.\delta_1$$

\noi where  $j>0$ is an integer,
$n_j > n_{j-1} > \ldots  > n_1 $  are integers $\geq 0$, and 
$\delta_j, \delta_{j-1}, \ldots , \delta_1$ are non null  ordinals $<\om_1$, i.e.  non null  countable ordinals.
In a second step it is proved in  \cite{Duparc03}, using infinite multi-player games,  that if such  an ordinal $\alpha$ is the Wadge degree of a  deterministic
 context-free $\om$-language, 
then all the ordinals $\delta_j, \delta_{j-1}, \ldots , \delta_1$ appearing in its Cantor normal form are smaller than the ordinal $<\om^\om$. Thus the Wadge 
hierarchy of the class $DCFL_\om$ is completely determined. 

\begin{The}[Duparc  \cite{Duparc03}]
The Wadge degrees of   {\it deterministic}      context-free $\om$-languages are exactly the ordinals in the form : 
$$\alpha = \om_1^{n_j}.\delta_j + \om_1^{n_{j-1}}.\delta_{j-1} + \ldots 
 + \om_1^{n_1}.\delta_1$$

\noi where  $j>0$ is an integer,
$n_j > n_{j-1} > \ldots  > n_1 $  are integers $\geq 0$, and 
$\delta_j, \delta_{j-1}, \ldots , \delta_1$ are non null  ordinals $<\om^\om$.
\nl The length of the Wadge hierarchy of the class $DCFL_\om$ is the ordinal $(\om^\om)^{\om} = \om^{(\om^2)}$. 
\end{The}

\noi 
Notice that theWadge hierarchy of $DCFL_\om$ is not determined in an effective way in  \cite{Duparc03}. The question of the decidability of problems like: 
``given two DMPDA $\mathcal{A}$ and $\mathcal{B}$, does $L(\mathcal{A}) \leq_W L(\mathcal{B})$ hold ?" or 
``given a DMPDA $\mathcal{A}$ can we compute $d^0_W (L(\mathcal{A}))$?" naturally arises. 
\nl Cohen and Gold proved that one can decide whether
an effectively given $\om$-language in  $DCFL_\om$  is an open or a closed set \cite{CG}.
 Linna characterized the \ol s accepted by DBPDA as the ${\bf \Pi}^0_2$-sets  in $ DCFL_\om$
 and proved in \cite{Linna77} that one  can decide whether an effectively given $\om$-language accepted by a DMPDA  is
 a ${\bf \Pi}^0_2$-set or a  ${\bf \Si}^0_2$-set.
\nl Using a recent result of Walukiewicz on  infinite games played on pushdown graphs,  \cite{Wal}, 
these decidability results were extended   in \cite{Fin01a} where it 
 was proved that one can decide whether a {\it deterministic}      context-free $\om$-language accepted by a given DMPDA is in a given 
  Borel class  ${\bf \Si}^0_1$, ${\bf \Pi}^0_1$, ${\bf \Si}^0_2$, or ${\bf \Pi}^0_2$ or even in the 
 wadge class $[L]$ given by   any \orl~ $L$.

\hs 
An effective extension of the Wagner hierarchy 
to  \ol s 
accepted by Muller {\it deterministic} real time   blind (i. e. without zero-test)  $1$-counter 
 automata has been determined in \cite{Fin01csl}.  
Recall that  blind  $1$-counter  automata form a subclass of $1$-counter automata hence also of pushdown automata. 
A  blind  $1$-counter Muller  automaton is just a  Muller pushdown automaton  $M=(K,X,\Gamma, \delta, q_0, Z_0, \mathcal{F})$
 such that $\Ga=\{Z_0, I\}$ where $Z_0$ is the 
bottom symbol and always remains at the bottom of the store. 
Moreover every transition 
which is enabled at zero level is also enabled at non zero level, i.e. if 
$\delta(q, a, Z_0)=(p, I^nZ_0)$, for some $p, q\in K$, $a\in X$ and $n\geq 0$, then 
$\delta(q, a, I)=(p, I^{n+1})$.  But the converse may not be true, i.e. 
some transition may be enabled at non zero level but not at zero level. Notice that  blind $1$-counter automata are sometimes called partially 
 blind $1$-counter automata as in \cite{Gre78}.
\nl  The Wadge  hierarchy of blind counter \ol s, accepted by deterministic Muller real time  blind  $1$-counter automata (MBCA), 
is studied in \cite{Fin01csl} in a similar way as Wagner studied the Wadge 
hierarchy of regular $\om$-languages in \cite{Wagner79}. 
Chains  and superchains for MBCA  are defined as Wagner did for Muller automata. The essential difference between 
the two hierarchies relies on the existence of superchains of transfinite  length $\alpha < \om^2$ for MBCA when in the case 
of Muller automata the superchains have only finite lengths. The hierarchy of $\om$-languages accepted by MBCA 
is effective and leads to effective winning strategies in Wadge games between two players in charge of $\om$-languages accepted by MBCA.
Concerning the length of the Wadge hierarchy of MBCA the following result is proved : 

\begin{The}[Finkel    \cite{Fin01csl}] \noi 
\begin{enumerate}
\ite[(a)] The length of the Wadge hierarchy of blind counter \ol s in  ${\bf \Delta^0_2}$
is $\om^2$.  
\ite[(b)] The length of the Wadge hierarchy of blind counter \ol s is the ordinal $\om^\om$ 
(hence it is equal to the length of the Wagner hierarchy).  
\end{enumerate}
\end{The}

\noi Notice that the  length of the Wadge hierarchy of blind counter \ol s is equal to the length of the Wagner hierarchy although it is actually 
a strict extension of  the Wagner hierarchy, as shown already in item (a) of the  above theorem. 
The Wadge degrees of  blind counter $\om$-languages are  the ordinals in the form : 
$$\alpha = \om_1^{n_j}.\delta_j + \om_1^{n_{j-1}}.\delta_{j-1} + \ldots 
 + \om_1^{n_1}.\delta_1$$

\noi where  $j>0$ is an integer,
$n_j > n_{j-1} > \ldots  > n_1 $  are integers $\geq 0$, and 
$\delta_j, \delta_{j-1}, \ldots , \delta_1$ are non null  ordinals $<\om^2$.
Recall that in the case of Muller automata, the ordinals $\delta_j, \delta_{j-1}, \ldots , \delta_1$ are non-negative integers, i.e. non null  ordinals $<\om$.

\hs Notice that Selivanov has recently determined the Wadge hierarchy of $\om$-languages 
accepted by {\it deterministic} Turing machines; 
its length is $(\om_1^{\mathrm{CK}})^\om$    
\cite{Selivanov03a,Selivanov03b}.  The $\om$-languages accepted by deterministic Muller Turing machines  or equivalently which are boolean combinations of 
arithmetical $\Pi^0_2$-sets have Wadge degrees in the form : 

$$\alpha = \om_1^{n_j}.\delta_j + \om_1^{n_{j-1}}.\delta_{j-1} + \ldots 
 + \om_1^{n_1}.\delta_1$$

\noi where  $j>0$ is an integer,
$n_j > n_{j-1} > \ldots  > n_1 $  are integers $\geq 0$, and 
$\delta_j, \delta_{j-1}, \ldots , \delta_1$ are non null  ordinals $<\om_1^{\mathrm{CK}}$.

\section{Topology and ambiguity in context-free $\om$-languages}\label{section-amb}

The notions of ambiguity and of degrees of ambiguity are well known and important in the study of context-free languages. 
These notions have been extended to context-free $\om$-languages accepted by B\"uchi or Muller pushdown automata in \cite{Fin03b}. 
Notice that it is proved in  \cite{Fin03b}  that these notions  are independent of the B\"uchi or Muller acceptance condition.  
So in the sequel we shall only  consider  the B\"uchi  acceptance condition. 

\hs We now firstly introduce a slight modification in the definition of a run of a B\"uchi pushdown automaton, which will be used in this section.

\begin{Deff}\label{run}
Let   $\mathcal{A}=(K,X,\Ga, \delta, q_0, Z_0, F)$ be a B\"uchi pushdown automaton. 
\nl
Let $\sigma =a_1a_2\ldots a_n\ldots$ be an $\om$-word over $X$. 
A run of $\mathcal{A}$ on $\sigma$ is an 
infinite sequence $r=(q_i,\gamma_i, \varepsilon_i)_{i\geq 1}$ where 
$(q_i,\gamma_i)_{i\geq 1}$ is an infinite sequence of configurations of $\mathcal{A}$
and,  for all $i \geq 1$,  $\varepsilon_i \in \{0, 1\}$ and: 

\begin{enumerate}
\ite $(q_1,\gamma_1)=(q_0, Z_0)$
\ite for each $i\geq 1$, there exists $b_i\in X\cup\{\lambda\}$ 
satisfying 
\nl $b_i: (q_i,\gamma_i)\mapsto_\mathcal{A}(q_{i+1},\gamma_{i+1} )$
\nl and ( $\varepsilon_i=0$ iff $b_i=\lambda$ ) 
\nl and  such that  ~ $a_1a_2\ldots a_n\ldots =b_1b_2\ldots b_n\ldots$ 
\end{enumerate}

\noi As before the \ol~ accepted by $\mathcal{A}$ is 
 $$L(\mathcal{A})= \{ \sigma\in X^\om \mid \mbox{  there exists a  run } r
\mbox{  of } \mathcal{A} \mbox{  on } 
\sigma \mbox{  such that } In(r) \cap F \neq\emptyset \}$$

\end{Deff}

\noi Notice that the numbers $\varepsilon_i \in \{0, 1\}$ are  introduced in the above definition 
in order to distinguish runs of a BPDA which go through 
the same infinite sequence of configurations but 
for which $\lambda$-transitions do not occur 
at the same steps of the computations.  

\hs As usual the  cardinal of $\om$ is denoted $\aleph_0$ and 
  the cardinal of the continuum is denoted $2^{\aleph_0}$. The latter is  
also the cardinal of the set of real numbers or of the set 
$X^\om$ for every finite alphabet $X$ having at least two letters.

\hs We are now ready to define degrees of ambiguity for BPDA and for context-free $\om$-languages. 

\begin{Deff} 
Let $\mathcal{A}$ be a BPDA reading infinite words over the alphabet $X$. 
For $x\in X^\om$ 
let $\alpha_\mathcal{A}(x)$ be the cardinal of the set of accepting runs of $\mathcal{A}$ on 
$x$. 
\end{Deff}

\begin{lem}[\cite{Fin03b}]\label{nb}
Let $\mathcal{A}$ be a BPDA  reading infinite words over the alphabet $X$. 
Then for all $x\in X^\om$ it holds that 
~~~~$\alpha_\mathcal{A}(x) \in \mathbb{N} \cup \{\aleph_0, 2^{\aleph_0}\}$.  
\end{lem}

\begin{Deff} 
Let $\mathcal{A}$ be a BPDA reading infinite words over the alphabet $X$. 
\begin{enumerate} 
\ite [(a)] If $\sup \{\alpha_\mathcal{A}(x) \mid  x\in X^\om \} \in \mathbb{N} \cup 
\{2^{\aleph_0}\}$, then 
$\alpha_\mathcal{A} = \sup \{\alpha_\mathcal{A}(x)   \mid   x\in X^\om \}$. 
\ite [(b)] If $\sup \{\alpha_\mathcal{A}(x)   \mid   x\in X^\om \} = \aleph_0$ and there 
is 
no word $x\in X^\om$ such that $\alpha_\mathcal{A}(x)=\aleph_0$, then 
$\alpha_\mathcal{A} = \aleph_0^-$. 
\nl ($\aleph_0^-$ does not represent a cardinal but is a new symbol that is 
introduced here to conveniently speak of this situation). 
\ite [(c)] If $\sup \{\alpha_\mathcal{A}(x)   \mid   x\in X^\om \} = \aleph_0$ and there 
exists (at least) 
one word $x\in X^\om$ such that $\alpha_\mathcal{A}(x)=\aleph_0$, then 
$\alpha_\mathcal{A} = \aleph_0$ 
\end{enumerate} 
\end{Deff}

\noi Notice that for a BPDA $\mathcal{A}$, $\alpha_\mathcal{A}=0$ iff 
$\mathcal{A}$ does not accept any 
$\om$-word. 
\nl We shall consider below that $\mathbb{N} \cup \{\aleph_0^-, \aleph_0, 2^{\aleph_0}\}$ 
is linearly ordered by the relation $<$, which is  defined by : 
$\forall k \in \mathbb{N}$, $k < k+1 < \aleph_0^- < \aleph_0 < 2^{\aleph_0}$.

\begin{Deff} For  $k \in \mathbb{N} \cup \{\aleph_0^-, \aleph_0, 
2^{\aleph_0}\}$  let 
\nl $CFL_\om(\alpha \leq k) = \{L(\mathcal{A}) \mid  \mathcal{A} 
\mbox{ is a } BPDA \mbox{ with } 
\alpha_\mathcal{A} \leq k \}$  
\nl $CFL_\om(\alpha < k) = \{L(\mathcal{A}) \mid  \mathcal{A} 
\mbox{ is a } BPDA \mbox{ with } 
\alpha_\mathcal{A} < k \}$  
\nl $NA-CFL_\om = CFL_\om(\alpha \leq 1)$ is the class of non ambiguous 
context-free $\om$-languages. 
\nl For every integer $k$ such that $k \geq 2$, or $k\in \{\aleph_0^-, 
\aleph_0, 2^{\aleph_0}\}$,
\nl $A(k)-CFL_\om = CFL_\om(\alpha \leq k) - CFL_\om(\alpha < k)$
\nl If $L \in A(k)-CFL_\om$ with $k \in \mathbb{N}, k \geq 2$, or $k\in \{\aleph_0^-, 
\aleph_0, 2^{\aleph_0}\}$, then $L$ is said to be inherently ambiguous 
of degree $k$. 
\end{Deff}

\noi Notice that one can define in a similar way the degree of ambiguity of 
a finitary context-free language. If $M$ is a pushdown automaton 
 accepting finite words by final states (or by final 
states and topmost stack letter) then $\alpha_M 
\in \mathbb{N}$ 
or $\alpha_M=\aleph_0^-$ or $\alpha_M=\aleph_0$. 
However every context-free language is accepted by a  pushdown automaton 
 $M$ with $\alpha_M \leq \aleph_0^-$, \cite{ABB96}. 
We  denote the class 
of non ambiguous context-free languages by $NA-CFL$ and the class of 
inherently ambiguous context-free languages by $A-CFL$. Then one can state the following result. 

\begin{The}[\cite{Fin03b}]
$$NA{\rm-}CFL_\om \subsetneq \om {\rm-}KC(NA{\rm-}CFL)$$
$$A{\rm-}CFL_\om \nsubseteq \om {\rm-}KC(A{\rm-}CFL)$$
\end{The}

\noi We now come to the study of links between topology and ambiguity in context-free $\om$-languages \cite{Fin03b,Fink-Sim}.

\noi Using a Theorem of Lusin and Novikov, and another theorem of descriptive set theory, see \cite[page 123]{Kechris94}, Simonnet 
proved the following strong result which shows that non-Borel  context-free $\om$-languages have a maximum degree of ambiguity.

\begin{The}[Simonnet \cite{Fink-Sim}]\label{mainthe}
Let $L(\mathcal{A})$ be a context-free $\om$-language accepted by a BPDA $\mathcal{A}$ such that $L(\mathcal{A})$ 
is an analytic but non Borel set. The set of $\om$-words, 
which have $2^{\aleph_0}$ accepting runs by $\mathcal{A}$, has cardinality $2^{\aleph_0}$. 
\end{The}

\noi On the other hand,  it turned out that, informally speaking,  the  operation $A \ra A^\sim$  conserves globally 
the degrees of ambiguity of infinitary context-free languages (which are unions of a finitary context-free language and of a  context-free $\om$-language). 
Then, starting from known examples of finitary  context-free languages of a given degree of ambiguity, are constructed in 
\cite{Fin03b} some  context-free $\om$-languages of any finite Borel rank and which are non-ambiguous or of 
any finite degree of ambiguity or of degree $\aleph_0^-$. 

\begin{The}\label{amb}
\noi 
\begin{enumerate}

\ite  For each non negative integer $n\geq 1$, there exist ${\bf\Si}_n^0$-complete 
non ambiguous context-free $\om$-languages $A_n$
and ${\bf \Pi}_n^0$-complete non ambiguous context-free $\om$-languages $B_n$.

\ite  Let $k$ be an integer $\geq 2$ or $k=\aleph_0^-$. Then 
for each integer $n\geq 1$, there exist ${\bf \Si}_n^0$-complete 
context-free $\om$-languages $E_n(k)$ and ${\bf \Pi}_n^0$-complete 
context-free $\om$-languages $F_n(k)$ which are in $A(k)-CFL_\om$, i.e. 
which are inherently ambiguous of degree $k$. 
\end{enumerate}
\end{The}

\noi Notice that the $\om$-languages $A_n$ and $B_n$ are simply those which were constructed in the proof of Theorem \ref{the-cf}. 
On the other hand it is easy to see that the BPDA accepting the 
context-free $\om$-language which is Borel of infinite rank, constructed in \cite{Fin03c} using an iteration of the operation $A \ra A^\sim$, has an  
infinite  degree of ambiguity. 
And  $1$-counter B\"uchi automata accepting  context-free $\om$-languages of any Borel rank of an effective analytic set, constructed via simulation of 
multicounter automata, may also have a great degree of ambiguity. So this left open some questions we shall detail  in the last section.  

\hs We indicate now a new result which follows easily from the proof of Theorem \ref{thewad} sketched in Section \ref{cf} above, see  \cite{Fin-mscs06}. 
Consider an $\om$-language $L $ accepted by a {\bf deterministic} Muller Turing machine or equivalently by a  {\bf deterministic}  $2$-counter Muller 
automaton. We get first an $\om$-language $\theta_S(L) \subseteq X^\om$ which has the same topological complexity (except for finite Wadge degrees), 
and which is accepted 
by a {\bf deterministic} real time  $8$-counter Muller automaton $\mathcal{A}$. 
\nl Then  one can  construct  from $\mathcal{A}$ a 
$1$-counter Muller automaton $\mathcal{B}$, reading words over the alphabet $X \cup\{A, B, 0\}$, 
 such that         $h(  L(\mathcal{A})   )    \cup h(X^{\om})^- =   L(\mathcal{B}) \cup h(X^{\om})^-$, 
where $h : X^\om \ra  (X \cup\{A, B, 0\})^\om$ is the mapping defined in Section \ref{cf}. 
Notice that the $1$-counter Muller automaton $\mathcal{B}$ which is constructed  is now 
also {\bf deterministic}. 
\nl On the other hand it is easy to see, from the decomposition given in \cite[Proof of Lemma 5.3]{Fin-mscs06}, 
 that the $\om$-language  $h(X^{\om})^-$ is accepted by a $1$-counter B\"uchi automaton 
which has degree of ambiguity 2
 and the $\om$-language $L(\mathcal{B}) $  is in  $NA-CFL_\om = CFL_\om(\alpha \leq 1)$  
because it is accepted by a {\bf deterministic} $1$-counter Muller automaton.
Then we can easily infer, using \cite[Theorem 5.16 (c)]{Fin03b} that the $\om$-language 
$h(  L(\mathcal{A})   )    \cup h(X^{\om})^- =   L(\mathcal{B}) \cup h(X^{\om})^-$
is in $CFL_\om(\alpha \leq 3)$. And this $\om$-language has the same complexity as $L(\mathcal{A})$ 
Thus we can state the following result. 

\begin{The}\label{wadge-amb}
\noi 
For each $\om$-language $L$ accepted by a {\bf deterministic} Muller Turing machine there is an $\om$-language $L' \in CFL_\om(\alpha \leq 3)$, 
accepted by a  $1$-counter Muller automaton $\mathcal{D}$ with $\alpha_{\mathcal{D}}\leq 3$, 
such that $L \equiv_W L'$. 

\end{The}

\section{$\om$-Powers of context-free languages}\label{powers}

\noi The $\omega$-powers of finitary languages  
are \ol s in the form $V^\om$, where $V$ is a finitary language over 
a finite  alphabet $X$.  They appear very naturally in the  characterization of the class 
$REG_\om$  of  \orl s (respectively,  of the class $CFL_\om$ of context-free \ol s) 
 as the $\om$-Kleene closure 
of the family $REG$ of regular finitary languages (respectively,   of the 
family $CF$ of context-free finitary languages) .  
\nl 
The question of the topological complexity 
of $\om$-powers  naturally arises and  was raised by 
Niwinski \cite{Niwinski90}, Simonnet \cite{Simonnet92},  and Staiger \cite{Staiger97b}. 

\hs An  $\om$-power of a finitary language is always an analytic set because 
it is either the continuous image of a compact set $\{0,1, \ldots ,n\}^\om$ for $n\geq 0$
or of the Baire space $\om^\om$. 
\nl 
The first example of finitary language $L$ such that $L^\om$ is analytic but not Borel, and even ${\bf \Si}_1^1$-complete, was obtained in  \cite{Fin03a}. 
Amazingly the language $L$ was very simple and even accepted by a $1$-counter automaton. 
It was obtained   via a coding of infinite labelled  binary trees. 

\hs We now give a simple  construction of this language $L$ 
using the notion of substitution which we now recall. 
 A {\it substitution}  is defined by a mapping 
$f: X \ra \mathcal{P}(\Ga^\star)$, where $X =\{a_1, \ldots ,a_n\}$  and $\Ga$ are two finite alphabets, 
$f: a_i \ra L_i$ where for all integers $i\in [1;n]$, $f(a_i)=L_i$ is a finitary language 
over the alphabet  $\Ga$.
\nl Now this mapping is extended in the usual manner to finite words:
$f(a_{i_1} \ldots a_{i_n})= L_{i_1} \ldots L_{i_n}$,   
and to finitary languages $L\subseteq X^\star$: 
$f(L)=\cup_{x\in L} f(x)$.  
\noi If for each integer $i\in [1;n]$ the language  $L_i$ does not 
contain the empty word, then the mapping $f$ may be extended to $\om$-words: ~~
$f(x(1)\ldots x(n)\ldots )= \{u_1\ldots u_n \ldots  \mid \fa i \geq 1 \quad u_i\in f(x(i))\}$ 
\nl and to \ol s $L\subseteq X^\om$ by setting  $f(L)=\cup_{x\in L} f(x)$.   

\hs 
Let now $X=\{0,1\}$ and  $d$ be a new letter not in $X$ and 
$$D=\{ u.d.v  \mid  u, v \in X^\star ~and~ ( |v|=2|u|)~~ or ~~( |v|=2|u|+1)~ 
\}$$ 
\noi $D\subseteq (X \cup \{d\})^\star$ is a context-free language accepted by a $1$-counter automaton. 
Let $g:X\ra \mathcal{P}((X \cup \{d\})^\star)$ be the substitution 
defined by 
$g(a)=a.D$. As $W=0^\star1$ is regular, 
$L=g(W)$ is a context-free language and it is accepted by   a $1$-counter automaton.   Moreover  one can prove that  $(g(W))^\om$  is 
${\bf \Si}^1_1$-complete, hence a  non Borel set. This is done by reducing to this $\om$-language a well-known example of 
${\bf \Si}^1_1$-complete set : the set of infinite  binary trees labelled in the alphabet $\{0,1\}$ which have an infinite branch in the 
${\bf \Pi}^0_2$-complete set $(0^\star.1)^\om$, see \cite{Fin03a} for more details.

\begin{Rem} The $\om$-language $(g(W))^\om$ is context-free. By Theorem \ref{mainthe}
 every BPDA accepting 
$(g(W))^\om$ has the maximum ambiguity and $(g(W))^\om \in A(2^{\aleph_0})-CFL_\om$.  On the other hand 
we can  prove that $g(W)$ is a non ambiguous context-free language. This is used in \cite{Fink-Sim} to prove   that neither unambiguity nor ambiguity 
of  context-free languages
are preserved under the operation $V \ra V^\om$.  

\end{Rem}

\noi Concerning Borel $\om$-powers, it has been  proved  in \cite{Fin01a} that 
 for each integer $n\geq 1$, there exist some 
$\om$-powers of  context-free languages 
which are ${\bf \Pi}_n^0$-complete Borel sets. 
These results were obtained by the use of 
 a new operation $V \ra V^\approx$ over $\om$-languages, which is a slight modification of the operation  $V \ra V^\sim$. 
The new operation $V \ra V^\approx$ preserves $\om$-powers and context-freeness. More precisely if  $V=W^\om$ for some context-free language 
$W$, then $V^\approx = T^\om$ for some context-free language $T$ which is obtained from $W$ by application of a given context-free substitution. 
And it follows easily from \cite{Duparc01} that if $V \subseteq X^\om$ is a 
${\bf \Pi}_n^0$-complete set, for some integer $n\geq 2$,  then $V^\approx$ is a ${\bf \Pi}_{n+1}^0$-complete set. Then, starting from the 
${\bf \Pi}^0_2$-complete set $(0^\star.1)^\om$, we get some ${\bf \Pi}_n^0$-complete $\om$-powers of context-free languages for each integer $n \geq 3$.

\hs An iteration of the operation $V \ra V^\approx$   was used in  \cite{Fin04-FI}  to prove that        there exists a finitary language $V$ 
such that $V^\om$ is a Borel set of infinite rank. The language $V$ was a simple recursive language but it was not context-free. Later,  
with a modification of the construction, using a coding of an infinity of erasers previously defined in  \cite{Fin03c},  Finkel and Duparc got a 
context-free language $V$ such that  $V^\om$ is a Borel set  above the class ${\bf \Delta}_\om^0$, \cite{Fin-Dup}. 

\hs  The question of the Borel hierarchy of  $\om$-powers of finitary languages has been solved very recently by Finkel and Lecomte in \cite{Fin-Lec}, where 
a very surprising result is proved, showing that actually $\om$-powers exhibit a great topological complexity. For every non-null countable ordinal 
$\alpha$ there exist some ${\bf \Si}^0_\alpha$-complete $\om$-powers and also some ${\bf \Pi}^0_\alpha$-complete $\om$-powers.
But the $\om$-powers constructed in \cite{Fin-Lec} are not $\om$-powers of context-free languages, except for the case of 
a  ${\bf \Si}^0_2$-complete set. Notice also that an example of a regular language $L$ such that $L^\om$ is  ${\bf \Si}^0_1$-complete was given 
by Simonnet in \cite{Simonnet92}, see also \cite{Lecomte-JSL} .

\section{Perspectives and open questions}

\noi We give below a list of some open questions which arise naturally. The problems listed here seem important for 
 a better comprehension of   context-free $\om$-languages   but the list is not exhaustive.  

\subsection{Effective results}

In the {\it non-deterministic} case, the Borel and Wadge hierarchies of context-free $\om$-langua \nl ges  are not effective, \cite{Fin01a,Fin03a,Fin01b}. 
This is not surprising since most decision problems 
on context-free languages are undecidable. 
On the other hand we can expect some decidability results in the case of {\it deterministic} context-free $\om$-languages. 
We have already cited some of them : we can decide whether a deterministic context-free $\om$-language is in a given Borel class or even 
in the Wadge class $[L]$ of a given regular $\om$-language $L$. 
The most challenging question in this area would be to find an effecive procedure to  determine  the Wadge degree of an $\om$-language in the class 
$DCFL_\om$. 
\nl Recall that the Wadge hierarchy of the class $DCFL_\om$ is determined in a non-effective way in \cite{Duparc03}. On the other hand the Wadge 
hierarchy of the class of blind counter $\om$-languages is determined in an effective way, using notions of chains and superchains, in \cite{Fin01csl}. 
There is a gap between the two hierarchies because (blind) $1$-counter automata are much less expressive than pushdown automata. One could try 
to extend the methods of \cite{Fin01csl} to the study of  {\it deterministic}  pushdown automata. 
\nl Another question concerns the complexity of decidable problems. A first question would be the following one. Could we 
extend  the results of Wilke and Yoo   to the class of blind counter $\om$-languages, i.e. 
is the   Wadge degree of  a blind counter $\om$-language computable in polynomial time ? Otherwise what is the complexity of this problem ? 
Of course the question may be further asked for classes of $\om$-languages which are located between the classes of blind counter $\om$-languages and 
of  deterministic context-free $\om$-languages. 
\nl Another interesting question would be to determine the Wadge hierarchy of $\om$-languages accepted by  deterministic higher order pushdown automata 
(even firstly in a non effective way), \cite{Engelfriet83,cw07report}. 

\subsection{Topology and ambiguity}

\noi Simonnet's Theorem \ref{mainthe} states  that non-Borel  context-free $\om$-languages have a maximum degree of ambiguity, i.e. are in the class 
$A(2^{\aleph_0})-CFL_\om$. 
On the other hand,   there exist some non-ambiguous context-free $\om$-languages of every finite Borel rank. 
The question naturally arises whether  there exist some non-ambiguous context-free $\om$-languages which are  Wadge equivalent to any given 
{\bf Borel}  context-free $\om$-language (or equivalently to any {\bf Borel}   $\Si_1^1$-set, by Theorem \ref{thebor}). 
This may be connected to a result  of Arnold who proved in \cite{Arnold83} that every Borel subset of 
$X^\om$, for a finite alphabet $X$,  is accepted by a {\it non-ambiguous} finitely branching  transition system with B\"uchi acceptance   
condition. 
By  Theorem \ref{amb},  if $k$ is an integer $\geq 2$ or $k=\aleph_0^-$, then 
for each integer $n\geq 1$, there exist ${\bf \Si}_n^0$-complete 
context-free $\om$-languages $E_n(k)$ and ${\bf \Pi}_n^0$-complete 
context-free $\om$-languages $F_n(k)$ which are in $A(k)-CFL_\om$, i.e. 
which are inherently ambiguous of degree $k$. 
More generally the question arises : determine  the Borel ranks and the Wadge degrees of context-free $\om$-languages in classes 
$CFL_\om(\alpha \leq k)$ or $A(k)-CFL_\om$ where $k \in \mathbb{N} \cup \{\aleph_0^-, \aleph_0, 
2^{\aleph_0}\}$ ( $k\geq 2$ in the case of $A(k)-CFL_\om$). 
A first result in this direction is Theorem \ref{wadge-amb} stated in Section \ref{section-amb}.

\subsection{$\om$-Powers}

\noi  The results of \cite{Fin01a,Fin03a,Fin04-FI,Fin-Lec}
 show that $\om$-powers of finitary languages  have actually a great topological complexity. 
Concerning $\om$-powers of context-free languages we do not know yet what  are all their infinite  Borel ranks.  
However  the results of  \cite{Fin-mscs06}  suggest that $\om$-powers of context-free languages or even of languages accepted by $1$-counter automata
exhibit  also a great topological complexity. 
\nl Indeed Theorem \ref{thebor} states that there are $\om$-languages accepted by B\"uchi $1$-counter automata of every Borel rank (and even of every 
Wadge degree) of an effective analytic set.  
On the other hand each  $\om$-language accepted by a B\"uchi $1$-counter automaton can be written as a finite union 
$L = \bigcup_{1\leq i\leq n} U_i.V_i^\om$, where for each integer $i$, $U_i$ and $V_i$ are finitary languages accepted by $1$-counter automata. 
Then  we can conjecture  that there exist some $\om$-powers of  languages accepted by  $1$-counter automata which have 
Borel ranks up to the ordinal  $\gamma_2^1$, although these languages are located at the very low level in the complexity hierarchy of finitary languages. 
\nl Recall that a finitary language $L$ is a code (respectively, an $\om$-code) if every word of $L^+$  (respectively, every $\om$-word of $L^\om$)
 has a unique decomposition in words of $L$, \cite{BerstelPerrin85}. 
 It is proved in \cite{Fink-Sim} that if $V$ is a context-free language such that $V^\om$ is a non Borel set then there are $2^{\aleph_0}$ $\om$-words of 
$V^\om$  which have $2^{\aleph_0}$ decompositions in words of $V$; in particular, $V$ is really not an $\om$-code although it is proved in \cite{Fink-Sim}
that $V$ may be a code (see the example V=g(W) given in Section \ref{powers}). The following question about {\bf Borel } $\om$-powers  now arises : 
 are there some  context-free codes (respectively,  $\om$-codes) $V$ such that 
$V^\om$ is ${\bf \Si}^0_\alpha$-complete  or  ${\bf \Pi}^0_\alpha$-complete for a given countable ordinal $\alpha < \gamma^1_2$ ?

\end{document}